\documentclass[10pt,preprint2]{aastex}
\usepackage{epsfig,graphicx}
\usepackage{natbib}
\bibpunct{(}{)}{;}{a}{}{,}    
\usepackage{amsmath} 
\newcommand{\vect}[1]{\boldsymbol{#1}} 
\usepackage{mathrsfs}
\usepackage{sidecap}

\begin{document}

\title{The D$_1$ enigma and its physical origin}
\shorttitle{The D$_1$ enigma and its physical origin}

\author{ J. O. Stenflo$^{1,2}$} 
\affil{$^1$Institute for Astronomy, ETH Zurich, CH-8093 Zurich, Switzerland }
\affil{$^2$Istituto Ricerche Solari Locarno, Via Patocchi, 6605 Locarno-Monti, Switzerland}

\email{stenflo@astro.phys.ethz.ch} 

\begin{abstract}
The D$_1$ enigma is an anomaly, which was first observed on
  the Sun as a symmetric polarization peak centered in the core of the
  sodium D$_1$ line that is expected to be intrinsically
  unpolarizable. To resolve this problem the underlying physics was
  later explored in the laboratory for D$_1$ scattering at   
  potassium vapor. The experiment showed that the scattering phase
  matrix element $P_{21}$ is positive while $P_{22}$ is negative,
  although standard quantum scattering theory predicts that both
  should be zero. This experimental contradiction is currently the
  main manifestation of the D$_1$ enigma. Subsequent theoretical
  studies showed that such  
  polarization effects may arise if scattering theory is extended to
  allow for interference effects due to level splittings of the ground state, in
  contrast to standard scattering theory, which only allows for
  interferences from level splittings of the intermediate
  state. Previous attempts to implement this idea had to 
  rely on heuristic arguments to allow modeling of the experimental
  data. In the present paper we develop a formulation of the 
  theory that can be self-consistently applied to quantum systems with
  any combination of electronic and nuclear spins. No statistical
  equilibrium or optical pumping is needed. The atom is assumed to be
  unpolarized at the beginning of each scattering event. The theory is
  capable of explaining both the phase matrix behavior of the
  laboratory data and the existence of a symmetric polarization peak
  in the core of the solar D$_1$ line. We also use it to predict the 
  polarization structures that we expect to see in a next-generation
  laboratory experiment with the rubidium isotopes $^{87}$Rb and
  $^{85}$Rb.  
\end{abstract}

\keywords{polarization  -- scattering -- techniques: spectroscopic -- Sun:
  atmosphere -- line: profiles -- methods: laboratory: atomic}

\maketitle

\section{Introduction}\label{sec:intro}
It has been widely believed that such a fundamental physical process
as the quantum scattering of visible light is well understood, because
the theory was developed and tested already in the 1920s during the
initial phase of the development of quantum mechanics. The experimental
tests at that time were however crude in comparison with the
possibilites offered by today's technology, and they were largely
discontinued around 1935, when the focus of the physics community
turned to other topics. 

A renewed interest in the basic physics of polarized light scattering
arose via solar physics through the discovery that the Sun's spectrum
is richly structured in linear polarization by coherent scattering
processes. This polarization, referred to as SS2 or the ``Second Solar
Spectrum'' \citep[cf.][]{stenflo-sk96,stenflo-sk97}, is a scattering
phenomenon that is not caused by external magnetic fields 
although it is modified by them through the Hanle effect. The wealth of unfamiliar
and apparently anomalous polarization phenomena in SS2 led to 
the recognition of the need for deepened explorations of the
theoretical foundations and for 
experimental tests of the theory with the precision that today's
technology allows. The prime enigma that confronted the theory was the
anomalous Na\,{\sc
  i} D$_1$ 5896\,\AA\ line, because standard quantum theory predicts
that it should be nearly unpolarizable, in contradiction with the solar
observations, which revealed it to possess polarization structure. 

The main question then was whether the D$_1$ enigma is a problem of
solar physics or of quantum physics (or both). While considerable
theoretical efforts have been invested in the modeling of the solar Na\,{\sc
  i} D$_1$ line \citep[e.g.][]{stenflo-landi98,stenflo-belluzzid115},
it was recognized that a laboratory experiment on polarized scattering
under controlled conditions was required to obtain an unequivocal
answer to the question whether there is something that has been 
overlooked in quantum scattering theory. The Sun's dynamic and
stratified optically thick atmosphere with tangled magnetic fields and
complicated geometry is a complex medium that is governed by
many unknown parameters, which makes it unsuited as a laboratory for 
definite tests of fundamental physical theories. One needs to
set up an experiment that cleanly isolates the physical
processes to be explored, without confusion from irrelevant
effects. 

\begin{figure}[t]
\centering
\resizebox{\hsize}{!}{\includegraphics{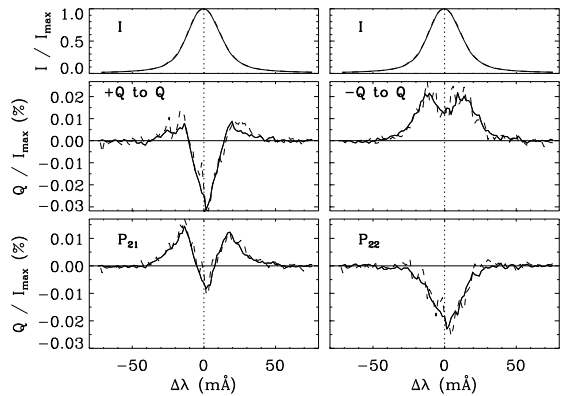}}
\caption{Laboratory results for scattering of linear polarization at
  potassium gas in the D$_1$ line at
7699\,\AA\ \citep[adapted from][]{stenflo-thalmannspw4}. The linear
polarization of the incident radiation is either perpendiculat ($+Q$)
or parallel ($-Q$) to the scattering plane, and Stokes $Q$ is measured
in the scattered radiation (middle panels, while the top panels give
the scattered Stokes $I$ or $P_{11}$). The phase matrix elements
$P_{21}$ and $P_{22}$ in the bottom panels are obtained from the two
middle panels as half the 
sum and half the difference. The circumstance
that the solid lines, which represent the use of full laser power, are
indistinguishable (within the noise fluctuations) from the dotted
  lines, which represent the case when the laser power has been
  reduced by a factor of 10, provides evidence that 
  we are in the regime of linear optics. Available scattering theory
  predicts zero polarization for the middle and bottom panels and is
  therefore unequivocally contradicted by the experiment. 
}\label{fig:SPW4_6}
\end{figure}

For this reason a laboratory experiment for polarized
D$_1$ scattering under the simplest possible controlled conditions
(single scattering at $90^\circ$) was carried out a decade
ago to determine the elements of the Mueller scattering matrix and
compare them with theoretical predictions, with the aim of either verifying or
falsifying available scattering theory. The experiment revealed
considerable 
polarization structure of the D$_1$ line where standard scattering
theory predicts null results
\citep{stenflo-thalmannspw4,stenflo-thalmannspw5}, which gave us an
unequivocal answer to our basic question: The D$_1$ enigma is indeed a
problem of quantum physics. Figure \ref{fig:SPW4_6} illustrates the
most relevant experimental results. 

The unavoidable conclusion is that any scattering theory that 
is unable to confront and explain the laboratory data must
be considered as incomplete. In the solar case the situation is more
ambiguous, because the Na\,{\sc
  i} D$_1$ polarization profile is observed to have a diversity of
shapes and amplitudes. However, among these various D$_1$ signatures it is
the existence of a symmetric D$_1$ polarization peak
\citep{stenflo-setal00a,stenflo-setal00b} that has remained enigmatic from the
point of view of quantum theory, and not so much the other more
anti-symmetric profiles. 

In the search for explanations of the enigmatic laboratory data it was
realized that if one would allow for interferences between pairs of scattering
amplitudes for which the initial magnetic substates or the final
substates of a given pair are
not the same (such terms are prohibited in standard scattering
theory), then the scattered radiation
could acquire polarization of the observed kind \citep{stenflo-s09}. These ideas
were further developed and used for modeling of the experimental data
in \citet{stenflo-s15costarica,stenflo-s15apj}. 

The difficulty was to show how these ideas
could be implemented in a self-consistent way. The phenomenological treatments in
\citet{stenflo-s09,stenflo-s15costarica,stenflo-s15apj} contained heuristic 
arguments that could not be
directly generalized for applications to arbitrary quantum systems. In the present
paper we replace this approach with a more well defined and consistent
theoretical framework, which provides an explanation for both the
laboratory results in Figure \ref{fig:SPW4_6} and for the existence of
a symmetric and positive polarization peak in the core of the solar Na
D$_1$ line. No optical
pumping or statistical equilibrium is used. The atom is assumed to be
unpolarized at the beginning of any given scattering event. It is
verified that the theory obeys the principle of spectroscopic
stability (in the limit of vanishing fine and hyperfine structure splitting)
for any combination of electronic and nuclear spins. 

A common reaction to the previous attempts to modify quantum
scattering theory has been one of disbelief, because
quantum mechanics has already survived scrutiny and experimental tests
for more than nine decades. This rejection is based on a
misunderstanding. The modifications that have been
proposed in the previous papers as well as in the present one do
not change anything at all in the Schr\"odinger equation, in the way we
calculate Feynman diagrams, or in the evolution of the wave function. The
computation of scattering probability amplitudes is not being
questioned. The problem lies exclusively in the procedure that we use to go
from the unobservable probability amplitudes to observables
(probabilities) in the case of multi-level atomic systems. This
procedure is not governed by any Schr\"odinger equation or Feynman
diagrams, and it has not been tested previously in the parameter
domain that is relevant to D$_1$ scattering. 

In a next-generation laboratory experiment we plan to test the theory with
collision-free scattering at rubidium vapor. We therefore show in Section
\ref{sec:outlook} the predicted linearly 
polarized D$_1$ profiles of the hyperfine structure components of the
two main isotopes $^{87}$Rb and $^{85}$Rb. Like for 
scattering at sodium and potassium the tell-tale signature of the new
scattering terms is that the D$_1$ phase
matrix elements $P_{21}$ and $P_{22}$ are non-zero and of opposite
signs. Without the new terms these matrix elements would be zero.

\section{Use of laser radiation to explore polarized scattering in the
weak radiation limit}\label{sec:laser}
Although the experimental falsification of standard scattering theory
as expressed by Figure \ref{fig:SPW4_6} is clear and unambiguous,
it has been dismissed on the grounds that laser light is special (because
of its coherent nature and high energy density) and cannot be used for
testing how natural radiation is scattered by physical
systems. Besides the remark that there cannot be anything wrong with
quantum mechanics this has been the unsubstantiated justification for
repeatedly rejecting papers on this topic. 

Let us therefore here address in some detail the question whether the
determination of the elements of the Mueller scattering matrix in any
way depends on whether a laser or any other kind of lamp is used as a
light source. There are two aspects of this: (i) the coherent nature
of laser light, and (ii) the high energy density of the laser beam, 
which makes the medium non-linear. In Figure \ref{fig:SPW4_6} the
indistinguishability (within the noise fluctuations) of the solid and
dashed curves, which represent recordings with full laser power
(solid) and the power reduced by a factor of ten (dashed), constitutes
experimental verification that there is no dependence of the
scattering polarization on the intensity of the incident
radiation. This implies that we are indeed in the regime of linear
optics. We will return to the issue of the energy density later in
this section, but with the experimental verification of no intensity
dependence, which implies that our results are representative of the
weak radiation limit, we next focus on the issue of the coherent
nature of laser radiation. 

As long as we are in the regime of linear optics, when the physical
properties of the medium to be studied do not depend on the ambient
radiation field, the polarizing properties of the medium are
completely described by the $4\times 4$ Mueller scattering matrix, which
relates the incoming and outgoing Stokes vectors to each other. When
dividing the matrix elements by the intensity to get the fractional polarizations, the
Mueller matrix contains 15 independent elements. They are determined
by sending in a sequence of light beams with known polarization states
(100\,\% linear, circular, etc.) and then measuring the Stokes vector
of the output beam. The medium can for instance be a weakly polarizing
instrument, a polarization modulation system, or a gas that polarizes
through light scattering. This medium can be considered as a ``black box'' with
unknown physical properties that are the sources of the polarization
effects. The measuring procedure to determine the 15 independent
Mueller matrix elements is independent of the internal physics of the
black box.

The coherence depth of the photons of the light source used for this
calibration of the Mueller matrix is irrelevant. If we determine the
polarization properties of an optical system, the results are
independent of the type of light source used, whether it is laser
radiation or a broad-band lamp (combined with a narrow-band filter, if
one wants good spectral selection). The only thing that matters is the
polarization state of the incident radiation. Because of the coherent
nature of laser light, it is naturally polarized, but by inserting
known polarization filters in the incident beam, we always ensure that all
the incident photons have identical polarization state, regardless of
whether the photons originate from a laser or from a lamp.

Natural light (like solar radiation) is only partially polarized,
because it consists of an ensemble of uncorrelated wave packets
(photons). When doing the ensemble averaging (incoherent summation)
one generally gets weak partial polarization, although each individual
component of the ensemble may represent 100\,\%\ elliptical
polarization. However, when letting natural (or laser) light pass
through a polarizing filter, one ensures that the polarization state
of all wave packets will be the same. The coherence length or relative
phase correlations between the wave packets are irrelevant for the
determination of the Mueller matrix, only the polarization state
matters.

We are free to mathematically decompose any light beam (natural or
laser) in its fundamental components, and then do ensemble averaging
over the bilinear products of their respective 
amplitudes. These bilinear products represent contributions to the
four Stokes parameters). The most natural 
mathematical decomposition is in terms of the Fourier components
(plane waves, infinite sine or cosine waves) of the radiation
field. We can then do the ensemble averaging over the bilinear
products of these
Fourier components. Nothing can be more coherent than a single Fourier
component (since it has infinite coherence depth). Laser radiation may
be thought of as an approximation of a single Fourier component per
polarization state. The ensemble average is then particularly simple,
but as mentioned before, this is irrelevant for our 
problem, because it is only the polarization state that matters, and
the polarization filter ensures that all the components of the
ensemble have identical polarization.

Let us now return to the often heard claim that laser radiation is
different because of its high energy density. In intense laser beams,
when the so-called Rabi frequency $\omega_R$ 
is much larger than the damping width, the atom gets ``dressed'' by
the radiation field \citep[cf.][]{stenflo-cohent77}, which leads to frequency
splitting and affects the positions, amplitudes, and widths of the
components in the scattered spectrum. $\hbar\omega_R$
represents the interaction energy between the atom and the radiation
field. The Rabi frequency depends on the radiation energy density of the beam
in the vapor cell, which in turn depends on the laser power and the cross
section of the expanded beam (in our case expanded to 1\,cm$^2$ to
fill the scattering region of the vapor cell). In Appendix \ref{sec:rabi} we have
computed the Rabi frequency for our experimental setup. When the laser
is used with maximum power (15\,mW) the Rabi frequency is 2.1 times
the radiative damping width, or 0.25\,m\AA\ in wavelength units. This
is well below the laser band width and about 38 times smaller than the
collisionally enlarged damping width. We are not aware of any theoretical
demonstration that this rather small value of the Rabi frequency would
cause any observable polarization effects (and we note that the level
splitting, being smaller than the laser band width, is not
observable). 

Still, the only conclusive way to rule out any possible non-linear
effects due to the energy density of the laser beam is to do it
experimentally. We have done the scattering measurements both 
with full laser power and with the power reduced by a factor of
ten. The results are found to be identical within the noise
fluctuations, as we have seen in Figure \ref{fig:SPW4_6}. This 
verifies that there is indeed no dependence on the
laser intensity over the covered range. This in turn
implies that the results are also valid in the weak
radiation limit. 

Doubters may argue that our factor of ten range does
not warrant extrapolation to the weak radiation limit. However, from
the constancy over a factor of 10 we have absolutely no reason to
expect that anything would change if we would further lower the laser
power by a factor of 100 or 1000. In principle such confirmation could
be carried out (and anyone who feels that it would be important could do this),
but since the noise level goes up when reducing the 
laser power, one would need to compensate with prolonged integration
times, until these times get prohibitively long. For the time being
the verification of the regime of linear optics over the factor of 10
is for all practical purposes good enough to allow us to proceed with
the quantitative modeling of the experiment on the assumption that we
are indeed in the regime of linear optics.

\subsection{Early attempts with ordinary lamps as light source}\label{sec:lamp}
It has repeatedly been argued that it would be better to use a normal laboratory
lamp as a light source for a scattering experiment. As a matter of fact this is what we did
when we began to develop the experiment more than a decade ago, but
this approach failed for two reasons: (1) The achievable S/N ratio was
much too small for detection of the subtle polarization effects in the D$_1$
line, and (2) it was not possible to sufficiently isolate D$_1$ from
possible contaminations caused by 
D$_2$. Also it was not feasible to spectrally resolve the various
components of the quantum system under study. 

Our first attempt was with a vapor cell for the yellow sodium lines, a
broad-band lamp as light source, and a Lyot element to discriminate
between the contributions from D$_1$ and D$_2$. This approach failed
because of insufficient discrimination blocking ratio due to the
finite acceptance angle of the Lyot element, in combination with the
wide solid angle of the scattered beam (needed to collect sufficient
photons). In addition the S/N ratio was insufficient to reveal the
subtle polarization effects in the D$_1$ line. 

We also considered spectral selection in the scattered beam. A
spectrograph solution can be ruled out because of vastly insufficient
optical throughput through its narrow entrance slit. One instead needs
a wide-angle narrow-band filter system. This insight led us to acquire two
tunable, lithium-niobate Fabry-Perot etalons that we could use in dual
configuration to suppress side lobes and ghosts. Still this approach
was found to be both cumbersome and not successful enough, because the
ghost suppression was generally insufficient to achieve the needed
discrimination ratio with respect to D$_2$. All this is documented in the
excellent PhD thesis of \citet{stenflo-fellerdiss}. It represents a
development phase that now is far behind us, and which led us to the
insight that a successful D$_1$ 
polarization experiment could only be done with a tunable laser, to
get the needed S/N ratio and spectral resolution. It was always clear
that one then needs to experimentally verify 
that we are in the regime of linear optics, and that when this
condition is satisfied the
coherent nature of laser radiation 
will not be an issue for the determination of the polarization
properties of the potassium gas.

\section{Coherency matrix formulation}\label{sec:cohmat}
The theoretical framework that has been developed in the monumental
monograph by \citet{stenflo-lanlan04} is based on the flat-spectrum
approximation (cf.~p.~257 of that monograph). This approximation requires 
that the incident radiation is broad-band, which implies that the
incoming wave packets have vanishing coherence depth. In contrast, we
know that actual 
natural light consists of wave packets with a coherence depths of
several meters, the distance that light travels during the time $1/\gamma$,
where $\gamma$ is the damping constant
\citep[cf.][]{stenflo-spw6s11}. 

It is a fundamental principle of quantum physics that any radiation
field (laser or natural light) can be described in terms of a linear
superposition of its Fourier components (or plane waves). We need a
scattering theory that allows us to compute the scattering probability
amplitude separately for each individual Fourier component (each of which has
infinite coherence depth). This is possible in the framework of the
Kramers-Heisenberg scattering formalism, but it is prohibited by the
flat-spectrum approximation.  

The Kramers-Heisenberg (K-H) dispersion formula for radiative
  scattering was introduced in the
  early days of quantum mechanics \citep{stenflo-kh25}, but
  \citet{stenflo-dirac47} later presented it in the notational form
  that we generally use today. The K-H scattering probability
  amplitude is the sum of two terms, one resonant term representing
  absorption followed by emission, and one non-resonant term
  representing emission followed by absorption. Here we can safely
  make the so-called rotating-wave approximation and 
  ignore the non-resonant term, since it is vanishingly small in comparison with the
  resonant one unless we are very far from the resonance. The K-H
  probability amplitude  for scattering $a\to b\to f$ is then 
\begin{equation}
w_{\alpha\beta}\sim \langle\,f\,\vert\,\hat{\vect{r}}\cdot
\vect{e}_\alpha\,\vert\,b\,\rangle\,\langle\,b\,\vert\,\hat{\vect{r}}\cdot
\vect{e}_\beta\,\vert\,a\,\rangle\,\Phi_{ba}\label{eq:scatkh}
\end{equation}
\citep[cf.][]{stenflo-book94,stenflo-s98}. 
Here 
\begin{equation}
\Phi_{ba}={2/i\over\omega_{ba}-\omega^\prime -i\gamma/2}
\,,\label{eq:phiba}
\end{equation}
where $\omega^\prime$ is the frequency of the incident radiation,
while 
\begin{equation}
\omega_{ba}=(E_b-E_{a})/\hbar
\label{eq:omba}\end{equation}
is the transition frequency between the energy levels of the
upper and lower magnetic substates $b$ and $a$. Indices $\alpha$ and
$\beta$ refer to the polarization states of the emitted and incident
radiation, respectively. 

Symbol $w_{\alpha\beta}$ represents the elements of a complex $2\times 2$
matrix $\vect{w}$. The way in which these elements are being calculated is not
questioned here. To go from these probability amplitudes, which are
unobservable, to probabilities that represent observable quantities,
we need to do ensemble averaging over the bilinear products
$w_{\alpha\beta}\,w_{\alpha^\prime\beta^\prime}^\ast$. They 
represent components of the $4\times 4$ coherency matrix $\vect{W}$
that is constructed from the $\vect{w}$ amplitude matrices via a tensor
product: 
\begin{equation}
\vect{W}\,=\,\Bigl(\,\sum_{abf} \,\vect{w}\,\Bigr)\otimes\Bigl(\,\sum_{a^\prime b^\prime f^\prime}
\,\vect{w}^\ast\,\Bigr)\,(\,\mathscr{S}_{\rm standard}\,+\,\mathscr{S}_{\rm new}).\label{eq:wmatnew} 
\end{equation}
This is an unfamiliar way of expressing $\vect{W}$, because we have
introduced a separate bracket with the two $\mathscr{S}$ terms, which
have the purpose of explicitly defining how the $a,b,f$ states should be combined
with the $a^\prime,b^\prime,f^\prime$ states. Far from all
combinations are allowed. The problem that we are addressing here is
how to correctly determine which combinations should be allowed, and
which should be prohibited. 

The standard way of writing $\vect{W}$ is without these $\mathscr{S}$
terms, while breaking up the summations into two categories: the
summations over the intermediate states $b$ and $b^\prime$, which are done
coherently (separately for $\vect{w}$ and $\vect{w}^\ast$), and the
summations over the initial and final states $a$ and $f$, which are done
incoherently (over the tensor products, not over the
amplitudes, with the condition that $a=a^\prime$ and $f=f^\prime$). 
The scattering process takes us from a definite initial to a definite final
state via all possible intermediate states. With this prescription the
only interference terms that are allowed are between 
scattering amplitudes that refer to different intermediate states $b$
and $b^\prime$. Interferences between scattering amplitudes for which
$a\ne a^\prime$ or $f\ne f^\prime$ are prohibited. 

According to the way in which Equation (\ref{eq:wmatnew}) has been
expressed, all summations are formally allowed to 
be coherent in the first two brackets. The mentioned restriction
that the initial and final states should be summed over incoherently is
enforced by the selection term $\mathscr{S}_{\rm standard}$. It 
can be expressed as the product of two Kronecker deltas, which
implement the selection rule $a=a^\prime$ and $f=f^\prime$: 
\begin{equation}
\mathscr{S}_{\rm standard} =\delta_{aa^\prime}\,\delta_{ff^\prime}\,.\label{eq:sstand} 
\end{equation}
In the present paper we will argue that the standard scattering scenario that
is defined by Equation (\ref{eq:sstand}) is too restrictive and
blocks the occurrence of valid interference terms that significantly affect the
scattering polarization that we can observe in the laboratory and on
the Sun. These previously overlooked terms can be formally unblocked
by adding the new ``selection term'' $\mathscr{S}_{\rm new}$ in
Equation (\ref{eq:wmatnew}). In Section \ref{sec:idnewterms} we address the
problem how to identify the missing terms and define the expression
for $\mathscr{S}_{\rm new}$ in a way that is 
unambiguous and physically
self-consistent for any quantum system with any combination of
electronic and nuclear spins. As a preparation and for conceptual
guidance we discuss in the next subsection how we can think about the
nature of the radiation-matter interaction process in the case of a
multi-level atomic system, and about the role of the tensor components of
the electric dipole moment.

\subsection{The resonant tensor components of the electric dipole
  moment}\label{sec:tensorcomp} 
The expression of Equation (\ref{eq:scatkh}) for the K-H scattering
amplitude contains two matrix elements, one for the absorption and one
for the emission leg of the scattering transition. Conceptually each
matrix element, e.g. the one that connects levels $b$ and $a$, can
be thought of as a resonant ``string'', in this example with resonant frequency 
$\omega_{ba}$ and ``string ends'' at states $\vert \,b\rangle$ and
$\vert a\rangle$. The string oscillations are excited by the electric
field - dipole interaction $\vect{d}\cdot\vect{E}^\prime$, where
$\vect{d}=-e \vect{r}$ is the electric dipole moment, and
$\vect{E}^\prime$ is the oscillating electric vector of the incident
radiation field. For a $J=0\to
1$ absorption transition we only sum over three ``strings'', which 
represent the three spatial components of $\vect{d}$. In this
particular case $\vect{d}$ can be treated as a vector. In the general
case, however, the dipole moment is a tensor, because the components
have two indices (representing the two ``end points''), and each such tensor
component represents a resonant string. 

\begin{figure}[t]
\centering
\resizebox{\hsize}{!}{\includegraphics{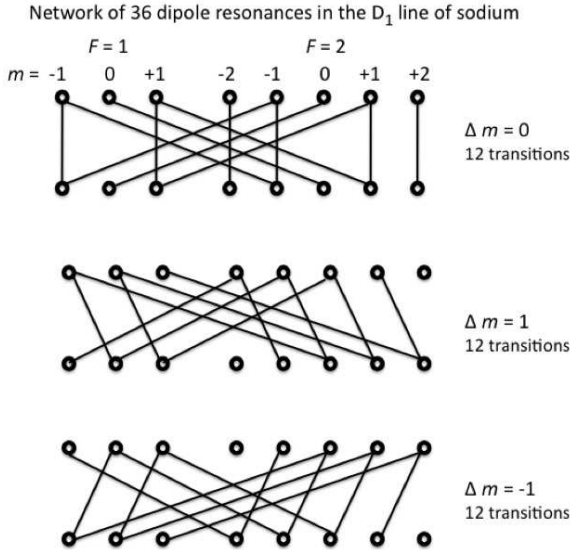}}
\caption{Illustration of the 36 resonances of the sodium D$_1$ line
  transition that are collectively excited by the
  driving incident radiation field. Each resonance represents a tensor
  component of the electric dipole moment of the atomic system. 
}\label{fig:res36}
\end{figure}

Figure \ref{fig:res36} illustrates that even for the relatively simple
case of the D$_1$ line of sodium the number of resonances or
``strings'' that collectively make up this $J=1/2\to 1/2$ transition
is quite large, namely 36. Potassium, barium, and lithium all have a
D$_1$ line with exactly the same resonance structure (although with
different resonance frequencies). All these chemical elements have nuclear
spin 3/2,  which induces a split of both the lower and upper $J=1/2$ states into
hyperfine structure states with total angular momentum quantum numbers
$F=1$ and 2. Since each $F$ state has $2F+1$ magnetic substates,
there are 8 substates each in the lower and upper levels. The number
of allowed electric dipole resonant transitions that connect them is
12 each for transitions with $q=\Delta m=0,\pm 1$, which makes a total
of 36 resonances. 

The electric-field component $E_q^\prime$ of the incident radiation simultaneously
drives the 12 oscillators with $m_a
-m_b=q$. In general the incident radiation field has 
contributions from all three $q$ values of $E_q^\prime$. Then 
the radiation field drives oscillations of all the 36 ``strings''. We focus
our discussion here on the D$_1$
line, because standard quantum scattering theory predicts it to be  a
polarization ``null line'', in unequivocal contradiction with
experimental data.

\subsection{Induced phase-locking between the scattering
  amplitudes}\label{sec:phaselock} 
In general all 36 resonators get driven by the oscillations of the
incident elctric field, and all these oscillators interfere with each
other as expressed by the tensor product in Equation
(\ref{eq:wmatnew}). However, unless the initially random phases between
a given pair of oscillators get phase-locked by the incident
radiation, the corresponding interference term vanishes when we do
ensemble averaging. 

In standard scattering theory there is an indirect way that has been
used to allow for
initial-state coherences, namely through preconditioning of the
atomic system by optical pumping that takes place prior to the examined scattering
event. Equation (\ref{eq:wmatnew}) can be generalized to include this
possibility, by attaching the lower level amplitude factors, $c_a$ attached to
$\vect{w}$, and $c_{a^\prime}^\ast$ attached to
$\vect{w}^\ast$. The preconditioning process has to include phase
synchronization between the respective sublevels, otherwise the
off-diagonal terms of the lower-state density matrix (that represents
ensemble averages over $c_a c_{a^\prime}^\ast$) would be zero. The
conclusion has been that if all off-diagonal terms of the ground-state
atomic density matrix are zero (at the beginning of the scattering
event), then there will be no contributions 
from scattering transitions with $a\ne a^\prime$, which leads to the
restriction that is expressed by Equation (\ref{eq:sstand}). 

The physics that we deal with here
is very different. No optical pumping is involved, and there is no
reference to any $c_a$ amplitude factors. They are not needed for the
same kind of reason that amplitude factors $c_b$ and $c_{b^\prime}$ are not
needed when we consider interference between the
intermediate states $b$ and $b^\prime$ in the sum over histories
scenario. At the beginning of the scattering event the off-diagonal
elements of the excited atomic state are zero (in the absence of
preconditioning by optical pumping). The phase synchronization between
states $b$ and $b^\prime$ 
that is a requirement for interference effects is accomplished by the 
driving electromagnetic field of the incident
radiation. Similarly, the incident
radiation is capable of synchronizing the
relative phases of initial states 
$a$ and $a^\prime$. Also this synchronization is done
without the need for any optical 
pumping or preconditioning of the atomic density matrix elements. 
In our present treatment the atomic system is instead assumed to be
unpolarized at the beginning of each given scattering event, i.e., all
off-diagonal density matrix elements are assumed to be zero. 

There is an additional reason why the atomic amplitude factors $c_a$
and $c_{a^\prime}$ or the corresponding density matrix elements should not
appear in our expression for the coherency 
matrix. Their presence would in general imply that the coherency
matrix (and therefore also the Mueller scattering matrix) is not just
a function of the properties of the material medium but also of the
ambient radiation field (its intensity, in combination with its
polarization state and directional 
properties). This means that we would be in the regime of  non-linear
optics. Here our aim is to formulate a theory that is representative
of the weak radiation limit of linear optics, and our experimental
tests with varying laser power verify that the experiment indeed deals with
that regime. With no reference to the atomic amplitude factors in the
expression of Equation (\ref{eq:wmatnew}) for the coherency matrix we
achieve that the scattering matrix is expressed in a way that is
manifestly independent of the ambient radiation field, and that it
therefore represents the regime of linear optics.

\section{Procedure to identify the new interference terms in an
  unambiguous way}\label{sec:idnewterms}
The tensor product in Equation (\ref{eq:wmatnew}) allows for a large
number of bilinear products between scattering probabilities, each of
which could potentially be a source of observable interference
effects, provided that they survive ensemble averaging. The
restriction that defines the subset of cross terms, which correspond
to physically valid interference terms, is represented by the selection
factor $\mathscr{S}$. The mathematical expression for this factor has
to be physically consistent for any quantum system. This consistency
requirement can be exploited to eliminate ambiguities in the
determination of $\mathscr{S}$ and can be expressed in terms of two
constraints that we will apply: (i) The Principle of Spectroscopic
Stability, and (ii) the condition for phase-locking by the radiation
field. Next we will elaborate on these two constraints and arrive at a
unique expression for $\mathscr{S}_{\rm new}$.

\subsection{Principle of spectroscopic stability as a selection
  criterion}\label{sec:pss}
In the absence of electron spin $S$ and nuclear spin $I$ ground states
cannot be split or be polarized, and the scattering transition is of
the ``classical'' type $L=0\to 1\to 0$. In the presence of fine
structure or hyperfine structure splitting, we get scattering
contributions from many more combinations of magnetic substates, as
indicated by Equation (\ref{eq:wmatnew}) and Figure
\ref{fig:res36}, with the potential for non-trivial selection terms
$\mathscr{S}$. A necessary (although not sufficient) requirement is then
that such terms must obey the Principle of Spectroscopic Stability
(PSS). 

The introduction of non-zero electronic and nuclear spins leads to
multiple, split atomic levels, and the various transition amplitudes
between the different levels get governed by complicated algebraic
expressions that depend on the new quantum numbers. All these
expressions contribute (with magnitude and sign) in an
intricate way when summing over the various scattering histories in
Equation (\ref{eq:wmatnew}) to compute the resulting scattering
matrix. PSS in this context means that if we let the splitting (both
fine and hyperfine) go to zero, such that the profile functions
$\Phi_{ba}$, which are part of the amplitudes $w_{\alpha\beta}$, will become
identical for all the transitions, while doing the full calculation of
the coherency matrix $\vect{W}$ with all the algebraic expressions for all
the allowed scattering transitions between the combinations of fine
and hyperfine structure states, then the resulting scattering matrix
should be identical to the classical one for an $L=0\to 1\to 0$
scattering transition. Letting the splitting go to
zero while retaining the non-zero values of spins $S$ and $I$ is
physically equivalent to replacing $S$ and $I$ by zero. While it is
clear that such a PSS must be obeyed to preserve physical consistency,
it looks like an ``algebraic miracle'' that it is also satisfied
mathematically, due to the apparent algebraic complexity of the
expressions over which we sum. The algebraic expression for the dipole
transition amplitude between two $m$ states is given in Appendix
\ref{sec:matelement}. 

Note that the selection term $\mathscr{S}$ in Equation
(\ref{eq:wmatnew}) has to obey the PSS criterion for all possible
combinations of electronic and nuclear spins $S$ and $I$. For instance,
in the case of a D$_1$ type transition $S=0.5$ while $I=1.5$. There
are a number of reasonable $\mathscr{S}$ term choices that satisfy PSS
for this particular combination of $S$ and $I$, but most of them fail to
satisfy PSS for certain other combinations. Such choices must be
rejected, because consistency requires that PSS must be satisfied for all possible $S$
and $I$ combinations. This criterion allows us to unambiguously weed out invalid
choices. The ``weeding''  has been done with a computer program that
loops through the various combinations of electronic and nuclear spins
to verify that the properties of the $L=0\to 1\to 0$ scattering matrix
are retrieved in the zero-splitting limit for all quantum-number combinations. 

In practice this is done by calculating the intrinsic polarizability
$W_2$ and requiring that for vanishing splitting it must become unity,
which is its classical dipole-scattering value. $W_2$ represents the
fraction of scattering processes that behave like classical,
Rayleigh-type dipole scattering, while the remaining fraction,
$1-W_2$, behaves like isotropic, unpolarized scattering. More
explicitly, the Mueller scattering matrix (also referred to as the phase
matrix), which describes how the Stokes vector is scattered, and which
is constructed from the coherency matrix $\vect{W}$ via a purely
mathematical transformation, can be written as \citep[cf.][]{stenflo-book94}
\begin{equation}
\vect{P}=W_2 \vect{P}_R\, +(1-W_2)\vect{E}_{11}\,+W_1 \vect{E}_{44}
\,{\textstyle{3\over 2}}\cos\phi\,,\label{eq:pw2} 
\end{equation}
where 
\begin{equation}
\vect{P}_R={3\over 4}
\begin{pmatrix}\,1+\cos^2\phi&\sin^2\phi&0&\,\,\,0\,\,\\ 
\,\sin^2\phi&1+\cos^2\phi&0&\,\,\,0\,\,\\ \,0&0&2\cos\phi&\,\,\,0\,\,\\ \,0&0&0&\,\,\,0\,\, \end{pmatrix}\label{eq:prmat} 
\end{equation}
is the classical Rayleigh phase matrix, $\phi$ is the scattering
angle, $\vect{E_{ii}}$ is a $4\times 4$ matrix with all elements zero
except for position $ii$, where it becomes unity, and $W_1$ is an intrinsic
polarizability coefficient that only relates to the circular
polarization. 

For any given fine-structure component, $W_2$ depends on the combination
of $J$ quantum numbers of the scattering transition. Let us take the
example of D$_2$ and D$_1$ when the nuclear spin in neglected. Then the
D$_2$ $J=0.5\to 1.5\to 0.5$ transition has $W_2=0.5$, while the D$_1$
$J=0.5\to 0.5\to 0.5$ transition has $W_2=0$. However, if one lets the
fine-structure splitting go to zero, so that the transition amplitude profiles of the
two lines superpose coherently, then $W_2$ becomes unity, as required by PSS. 

Using Equations (\ref{eq:pw2}) and (\ref{eq:prmat}) we can
express $W_2$ in terms of the matrix elements $P_{ij}$ in a way that
is independent of the scattering angle $\phi$, as follows: 
\begin{equation}
W_2 = {2\,(P_{12}+P_{22})\over 3\,P_{11}+2\,P_{12}-P_{22}}\,.\label{eq:w2fromp} 
\end{equation}
In our computer application of the PSS criterion we calculate $W_2$
according to Equation (\ref{eq:w2fromp}) 
from the phase matrix derived for a given combination of electronic and
nuclear spins, to check whether it satisfies the requirement of being
unity for all combinations of quantum numbers.

\subsection{Requirement of phase locking as a selection
  criterion}\label{sec:phlock}
As explained in Section \ref{sec:tensorcomp}, the dipole transitions
that connect an upper and a lower $m$ state may 
be thought of as resonant  
strings with two ends, one in each connected $m$ state. The string 
oscillations are driven by the radiation field. When forming the
coherency matrix $\vect{W}$ in Equation (\ref{eq:wmatnew}) we get
products between pairs of oscillators. These products will
not survive ensemble averaging if the relative phases of the two
oscillators are random. If however the two
strings have one common end point, then the oscillations   
will get synchronized by the radiation field. If on the other hand neither the upper nor the
lower pair of end points coincide for a given pair of strings, then
the relative phase is random and the ensemble average vanishes. This
random phase relation follows from 
our assumption that there is no atomic polarization at the start of the
interaction with the radiation field, which implies that states $a$ and
$a^\prime$ are uncorrelated when they refer to different $m$ states. 

In standard scattering theory each string pair has common 
lower ends ($a=a^\prime$ and $f=f^\prime$), while the upper ends may
differ, as expressed by the selection term $\mathscr{S}_{\rm
  standard}$ in Equation (\ref{eq:sstand}). It is readily verified that this term obeys
PSS for all combinations of electronic and nuclear spins. 

This is however not the only way to enable phase locking by the
radiation field. It would become possible if we tie together the upper
end points instead of the lower string end points. This implies that
the new selection term $\mathscr{S}_{\rm 
  new}$ should enforce $b=b^\prime$ and therefore contain the factor
$\delta_{bb^\prime}$. In addition we need to have either $a\ne
a^\prime$ or $f\ne f^\prime$ or both, because the special case when
all $a$, $b$, and $f$ ends are tied together is already included as a
special case of $\mathscr{S}_{\rm  standard}$. Therefore $\mathscr{S}_{\rm
  new}$ must also contain either the factor $1-\delta_{aa^\prime}$
or $1-\delta_{ff^\prime}$ or both. 

\begin{figure}[t]
\resizebox{\hsize}{!}{\includegraphics{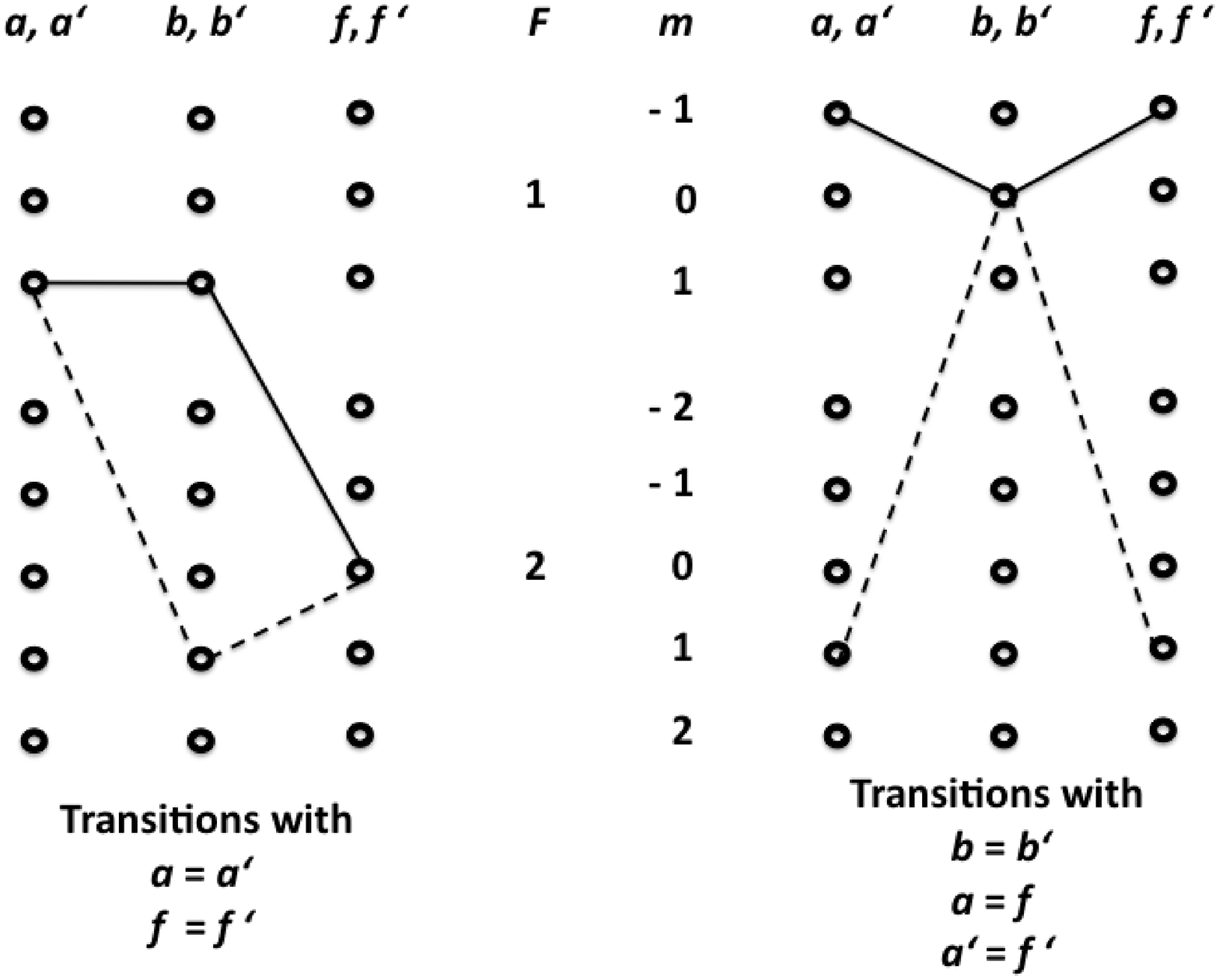}}
\caption{Illustration of D$_1$ scattering terms in the coherency matrix
  $\vect{W}$. The solid lines represent the amplitudes $\vect{w}$ for
  scattering $a\to b\to f$, while the dashed lines represent the
  amplitudes $\vect{w}^\ast$ for scattering $a^\prime\to b^\prime\to
  f^\prime$. The D$_1$ transition has 8 lower and 8 upper $m$ states,
  which may be connected by the scattering process. The left part of
  the figure illustrates the types of terms that are allowed  in standard scattering theory,
  as defined by parameter $\mathscr{S}_{\rm
      standard}$ in Equation (\ref{eq:sstand}), while the right part
  of the figure illustrates the types of new terms that are governed by parameter
  $\mathscr{S}_{\rm new}$ in Equation (\ref{eq:snew}). 
}\label{fig:scatdiag}
\end{figure}

\subsection{Using the combined constraints to obtain a unique
  expression for $\mathscr{S}_{\rm new}$}\label{sec:combconst}
When testing the different versions of $\mathscr{S}_{\rm
  new}$ that satisfy these phase-locking requirements we find that
only two versions also satisfy PSS, and that only the
following one of these two is also compatible with observations: 
\begin{equation}
\mathscr{S}_{\rm new}
  =(1-\delta_{aa^\prime})\,(1-\delta_{ff^\prime})\,\delta_{b\,b{\,\prime}}\,\,\delta_{af}\,\,\delta_{a^\prime
    f^\prime}\,.\label{eq:snew} 
\end{equation}
Note that in this expression we have left out a decoherence factor,
which appears when the level splitting is non-zero. The origin of this
decoherence factor will be dealt with in the next subsection. 

First we recognize that the needed essential property of this version of $\mathscr{S}_{\rm
  new}$ is that it contains the 
factors $\delta_{af}$ and $\delta_{a^\prime f^\prime}$ which enforce
that the initial and final $m$ states are the same for each scattering
amplitude of an interfering pair, as illustrated more explicitly in
the right portion of Figure \ref{fig:scatdiag}. This requirement makes
the scattering process behave mathematically as if
it would go from initial state $b$ back to the same state via the
intermediate states $a$ and $a^\prime$, although in the physical reality
it is $a$ and $a^\prime$ that are the initial states. 

The other version that also satisfies PSS is not unreasonable but a
bit contrived. It is obtained if 
$\delta_{af}\delta_{a^\prime f^\prime}$ in Equation (\ref{eq:snew})
is replaced by $\delta_{m_a m_f}\delta_{m_{a^\prime} m_{f^\prime}}$, i.e.,
the initial and final $m$ states are enforced to be identical for each
given scattering amplitude, while the initial and final $F$ states are
allowed to differ. However, this version has to be rejected
because it leads
to polarization effects that are not compatible with observations, in
particular with those from the laboratory experiment on K D$_1$ scattering. This
leaves us with Equation (\ref{eq:snew}) as the only remaining
physically consistent choice.

\subsection{Applying the $\mathscr{S}_{\rm new}$  selection to split
  multiplets}\label{sec:nonzerosplit} 
Having settled for this definite choice of selection factor
$\mathscr{S} $, which uniquely defines the subset of bilinear products
that contribute to the coherency matrix $\vect{W}$ in the
limit of vanishing splitting, let us next turn
to the general case when the 
fine or hyperfine structure splitting is non-zero. Our task is to
define how the interference and decoherence effects depend on the
non-zero splitting. 

For both the standard and new terms interference originates from the cross products
$\Phi_{ba}\Phi_{b^\prime a^\prime}^\ast$ that appear when we form the
bilinear products between scattering amplitudes $\vect{w}$ and
$\vect{w}^\ast$ in the expression for the 
coherency matrix $\vect{W}$ (cf. Equations
(\ref{eq:scatkh})--(\ref{eq:wmatnew})). The decoherence effects can be
factorized out 
from the frequency-dependent part by
converting the profile product to a profile sum \citep[cf.][]{stenflo-book94}:  
\begin{equation}
\Phi_{ba}\Phi_{b^\prime a^\prime}^\ast\sim \,\cos\beta\,\,
e^{i\beta}\,\,\textstyle{1\over 2}(\Phi_{ba}+\Phi_{b^\prime
  a^\prime}^\ast)\,,\label{eq:prod2sum}  
\end{equation}
where 
\begin{equation}
\tan\beta ={\omega_{b^\prime a^\prime}-\omega_{ba}\over \gamma}\,,\label{eq:tanbeta}  
\end{equation}
and $\gamma$ is the damping constant. The factor ${1\over 2}$ is
introduced to make the wavelength-dependent part of this expression
area normalized. The decoherence is described by
$\cos\beta\,\,e^{i\beta}$.  When angle $\beta =0$, the wavelength
integration over the right-hand side gives unity. 

When the interference involves only states with different $m$ quantum
numbers (for given $J$ and $F$ numbers), the decoherence is an
expression of the Hanle effect, and 
angle $\beta$ is referred to as the ``Hanle angle''. Since our
treatment is much more general and includes interferences between
states with any combination of quantum numbers (differing in $m$, $J$,
or $F$ or all of them at the same time), we will use the more general
term ``decoherence angle'' for $\beta$. 

In standard scattering theory $a=a^\prime$, so that
$\tan\beta=\omega_{b^\prime b}/\gamma$. Then all decoherence effects are
caused by splitting of the intermediate, excited state. For
the new terms that are defined by $\mathscr{S}_{\rm new}$ we instead
have $b=b^\prime$, and then interference only 
happens because $a\ne a^\prime$. In this case the decoherence angle is
determined by 
\begin{equation}
\tan\beta_g ={\omega_{a a^\prime}\over \gamma}\,,\label{eq:betag}  
\end{equation}
where $\omega_{a a^\prime}=(E_a -E_{a^\prime})/\hbar$ according to
Equation (\ref{eq:omba}). For clarity we have here introduced
subscript $g$ to mark that the decoherence is caused by the ground-state
splitting. 

In contrast to standard scattering theory the new terms need an
additional decoherence factor, which we here refer to as $k_{\rm
  decoh}$ to distinguish it from the rest. It originates from a beat
frequency between two separated intermediate metalevels within the
excited state $b$. Let us explain. For simplicity we assume that the
ground state has an infinite life time, which implies that the
substates $a$ and $a^\prime$ are infinitely sharp. If the incident
radiation has frequency $\omega^\prime$, then the radiatively excited
metalevel, which belongs to level $b$, will have an energy that is
$E_a +\hbar\,\omega^\prime$ in the case of the
scattering amplitude $\vect{w}$, while the intermediate metalevel for
the $\vect{w}^\ast$ scattering amplitude will have the different
energy $E_{a^\prime} +\hbar\,\omega^\prime$. 

The energy difference between the two metalevels gives rise to  
beat oscillations, which are exponentially damped (with damping
constant $\gamma$) because the excited state $b$ has a limited life
time. When doing ensemble averaging (in this case in the 
form of temporal integration over the damped oscillations) we get the
decoherence factor  
\begin{equation}
k_{\rm decoh}=\cos\beta_g\,e^{i\,\beta_g}\,,\label{eq:kdecoh}  
\end{equation}
which is identical to the one obtained from the
product of the profile functions and is also governed by Equation (\ref{eq:betag}). 
Combining the two factors
the total decoherence is found to be $\cos^2\beta_g\,e^{2i\,\beta_g}$. 

The appearance of two decoherence factors that may be combined this way is
familiar to us from PRD (partial frequency redistribution) theory: For
the frequency coherent term 
$R_{\rm II}$ decoherence only appears once, due to the profile products
of the absorption legs, while for the CRD term $R_{\rm III}$ there is
an additional factor from the emission process. The origin of the
second decoherence factor is however different in our
case. Since we are here only dealing with the collision-free case, the
scattering process is frequency coherent ($R_{\rm II}$ only) because
of energy conservation. To understand that there is no contradiction
it may be conceptually helpful to think of $k_{\rm decoh}$ as a branching ratio, because
the beat oscillations, which cause it, effectively reduce the life
time of the excited state. 

The reason why there is no corresponding decoherence factor for
``standard'' scattering as represented by $\mathscr{S}_{\rm standard}$ is that the energies of
the intermediate metalevels of the two scattering amplitudes are
identical, thus without any beat oscillations. Since in ``standard''
theory $a^\prime =a$, and level $a$ is assumed to be infinitely sharp,
the intermediate metalevel has energy $E_a +\hbar\,\omega^\prime$ for both
scattering amplitudes. 

All our discussions that have led to the identification of the new
interference terms and to their decoherence behavior may be expressed
in final form in terms of our selection factor
$\mathscr{S}=\mathscr{S}_{\rm standard} +\mathscr{S}_{\rm new}$ as follows: 
\begin{eqnarray}\label{eq:sscript}
&\phantom{=}&\mathscr{S}=\,\delta_{aa^\prime}\,\delta_{ff^\prime}\\
&\phantom{=}& +\, k_{\rm  decoh}\,\,(1-\delta_{aa^\prime})\,(1-\delta_{ff^\prime})\,\delta_{b\,b^{\,\prime}}\,\,\delta_{af}\,\,\delta_{a^\prime  f^\prime}\,.\nonumber  
\end{eqnarray}
Note that we have here attached the extra decoherence factor $k_{\rm
  decoh}$  to the second term on the right-hand side. For the first
term that represents standard theory this factor is unity and
therefore does not appear, for the reasons that we have just
explained. When this expression for $\mathscr{S}$ is used in Equation
(\ref{eq:wmatnew}), the bilinear products between the profile factors
$\Phi_{ba}$ automatically take care of the remaining decoherence
effects for both the standard and new terms. 

Let us finally make some clarifying remarks concerning the physical
origin of the damping constant $\gamma$, which governs the magnitude
of the interference and decoherence effects for the new terms of the
extended theory. In previous papers we have referred to the new
effects as due to ground-state coherences, but we now feel that such a
terminology is misleading and should be abandoned, because it
incorrectly suggests that the ground states have finite energy widths that cause
them to partially overlap, and that this is the origin of the
coherence effects. However, in our treatment we assume that the ground
states have infinite life time and are therefore infinitely
sharp. This implies that there is no partial overlap between their
energy levels. The damping that is the cause of the partial overlap,
decoherence effects, and interferences between the scattering
amplitudes has its origin exclusively in the excited, intermediate
state, both for the standard and the new scattering terms. Although
the new interference effects are governed by the splitting of the
infinitely narrow ground states, the damping that determines the
magnitude of the effects comes from the excited state. 

\begin{figure}[t]
\resizebox{\hsize}{!}{\includegraphics{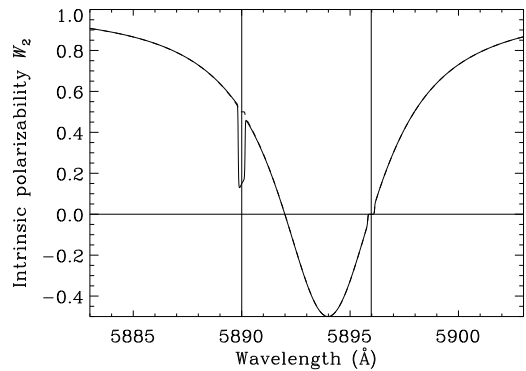}}
\caption{Intrinsic polarizability $W_2$ for the Na D$_2$ and D$_1$
  line system as computed from the scattering
  matrix elements via Equation (\ref{eq:w2fromp}). The $W_2$ curve
  obtained with the new theory as defined by Equation
  (\ref{eq:sscript}) is identical to the one obtained with
  standard theory. The trough around the D$_2$ resonance is caused by the
  hyperfine structure splitting. When this splitting is neglected, we
  get the dashed curve, which differs from the solid one only through
  the absence of a D$_2$ trough. 
}\label{fig:naw2}
\end{figure}

\subsection{Implications for the scattering matrix}\label{sec:implications} 
Applying our extended theory as defined by Equation (\ref{eq:sscript}) we
show in Figure \ref{fig:naw2} the 
wavelength variation of the intrinsic polarizability $W_2$ across the
Na D$_2$ and D$_1$ line system. Here $W_2$ has been derived from the phase
matrix elements through Equation (\ref{eq:w2fromp}). We find that we
get the identical $W_2$ function (the solid curve) with and without the
$\mathscr{S}_{\rm new}$ term in Equation (\ref{eq:wmatnew}). The new
interference effects thus do not affect $W_2$ for
the D$_2$ -- D$_1$ system. 

The trough-like depression of $W_2$ around the D$_2$ resonance is
induced by the hyperfine structure splitting. When we remove this
splitting, i.e., set $I=0$, we obtain the dashed curve in Figure
\ref{fig:naw2}, which coincides with the solid curve everywhere except
in the Doppler core of the D$_2$ line, where the $W_2$ trough
disappears and $W_2=0.5$. In the Doppler core of the D$_1$ line we
always get $W_2=0$. 

Note in Figure \ref{fig:naw2} that the $W_2$ curve asymptotically approaches unity in the far
line wings. When we get far from the
resonances in comparison with the magnitude of the fine-structure splitting, then the
effects of the splitting become small and the scattering behavior
approaches that of a classical dipole. 

\begin{figure}[t]
\resizebox{\hsize}{!}{\includegraphics{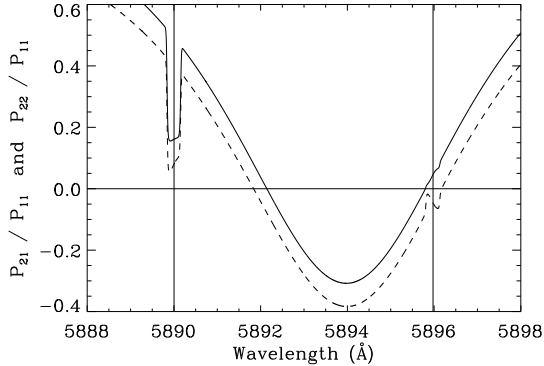}}
\caption{Symmetry breaking in the scattering matrix, illustrated for
  the $90^\circ$ scattering case. The solid curve represents the fractional polarization
  $P_{21}/P_{11}$, which is the linear polarization that one would get
  when the incident radiation is unpolarized. For comparison, the dashed
  curve represents the ratio $P_{22}/P_{11}$. While $P_{21}$ and $P_{22}$
  are the same in the standard theory, the new interference terms
  break this symmetry and cause the split between the solid and
  dashed curves. Notice in particular that at the D$_1$ resonance
  $P_{21}$ is positive, while $P_{22}$ is negative by the same
  amount. 
}\label{fig:p21p22}
\end{figure}

The new interferences instead manifest themselves in the form of a symmetry
breaking between the $P_{21}$ and $P_{22}$ components of the phase
matrix $\vect{P}$, as illustrated in Figure \ref{fig:p21p22} for the
$90^\circ$ scattering case. We
have plotted these components in normalized fractional polarization
form, $P_{21}/P_{11}$ as the solid line, and $P_{22}/P_{11}$ as the
dashed line. In standard scattering theory the solid and dashed lines
would coincide and go through zero at the D$_1$ resonance. The
systematic difference between them is exclusively caused by the new
interferences.  

Note that the relative shift between the curves is such that at the D$_1$
resonance $P_{21}/P_{11}$ is significantly positive, while
$P_{22}/P_{11}$ is negative by the same amount, in qualitative
agreement with the laboratory results for scattering in the
potassium D$_1$ line that were illustrated in Figure
\ref{fig:SPW4_6}. In Section \ref{sec:kd1} we will apply our new
theory to model the laboratory measurements in considerable
quantitative detail. First we will however apply the theory to explain
the enigmatic existence of a symmetric polarization peak centered in
the core of the solar Na D$_1$ line.

\section{Solar polarization of the Na\,{\sc i} D$_1$ and D$_2$ line
  system}\label{sec:nad1d2} 
Near (but inside) the solar limb the scattering geometry resembles
that of $90^\circ$ scattering, although the origin of the scattering
polarization is the small anisotropy of the incident radiation field
that manifests itself as the limb darkening of the solar disk. In 
good approximation while avoiding detailed angular integrations, one can
think of the anisotropic radiation field as consisting of two
components, one directional component that is the source of the
polarization, and one isotropic component that dilutes the polarized
radiation with unpolarized light. This dilution factor enters as a
global scaling factor if we ignore the wavelength dependence of the
anisotropy across the considered spectral window. 

For exploratory, initial modeling of polarized line profiles before going
into the technical complexities of polarized radiative transfer with
partial frequency redistribution, it has often been found very useful
to apply the Last Scattering Approximation \citep[LSA,
cf.][]{stenflo-s80,stenflo-book94}. In the LSA 
scenario the last scattering event occurs at a particle that is
illuminated by unpolarized incident radiation. The fractional linear
polarization $Q/I$ of the scattered radiation will then be
$P_{21}/P_{11}$, if one ignores the geometric dilution factor and the
superposed contributions from continuum radiation. 

The continuum introduces two effects: it dilutes the line opacity with
a spectrally flat opacity, and it adds a small, spectrally flat
polarization contribution. With these ingredients we can construct a
very simple but still insightful model of the scattering polarization
across the Na\,{\sc i} D$_2$ -- D$_1$ system, as follows: 
\begin{equation}
(Q/ I)_{\rm model}(\%)=s\,\,\,{P_{21}+a_c\,b_c\over P_{11}+a_c}\,.\label{eq:namodel}
\end{equation}
While $P_{21}$ and $P_{11}$ are elements of the phase matrix for
scattering in the sodium resonance lines, the wavelength-independent
parameters $a_c$ and $b_c$ represent the spectrally flat continuum
opacity and polarization, respectively. Parameter $s$ is a global
scaling factor that accounts for the geometric dilution effects. A
similar model was introduced in \citet{stenflo-s80} 
to successfully model the observed polarization across the Ca\,{\sc
  ii} K and H lines with the spectacular signature of quantum
interference between the two total angular momentum states. 

If there were no continuum opacity, $Q/I$ would be $s\, P_{21}/P_{11}$
and look like the solid curve in Figure \ref{fig:p21p22}, 
only differing in the scaling factor, which has no effect on the
shape of the curve. In the far wings both $P_{21}$ and $P_{11}$ go to
zero, with the consequence that $Q/I$ asymptotically approaches
$s\,b_c$, the continuum polarization level. 

\begin{figure}[t]
\resizebox{\hsize}{!}{\includegraphics{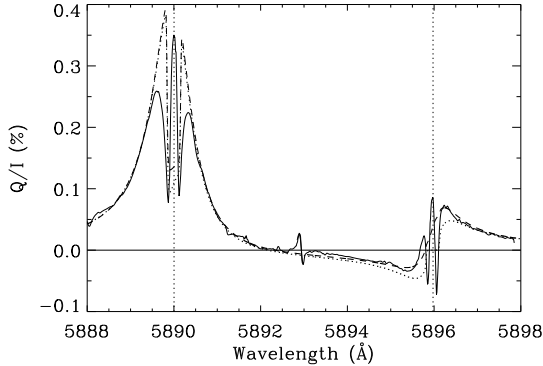}}
\caption{Modeling of the observed  (solid curve) linear polarization $Q/I$ across the solar Na
  D$_2$ and D$_1$ lines, obtained from the observations of
  \citet{stenflo-setal00a} with the spectrograph slit 5\,arcsec inside
  the solar  limb. The dashed curve is based on the new theory, 
  the dotted curve on the standard 
  theory. The model uses LSA (last scattering approximation) as
  expressed by Equation (\ref{eq:namodel}). Notice in
  particular that the anti-symmetric behavior of the dotted model
  curve around the D$_1$ resonance gets broken by the new interference
  terms, which gives rise to a systematic positive polarization surplus
  centered around the D$_1$ resonance wavelength. 
}\label{fig:namodel}
\end{figure}

The most detailed, high S/N ratio explorations of the non-magnetic
scattering polarization across the Na\,{\sc i} D$_2$ -- D$_1$ system
were carried out with the ZIMPOL imaging polarimeter at the NSO/Kitt
Peak McMath-Pierce facility in 1998 \citep{stenflo-setal00a}. The solid
curve in Figure \ref{fig:namodel} shows the observed fractional $Q/I$
polarization recorded with the spectrograph slit 5\,arcsec inside the
solar limb. It shows a remarkable polarization peak not only in the
Doppler core of the D$_2$ line, but at the center of the D$_1$ line as
well, where it was totally unexpected, because the intrinsic
polarizability $W_2$ is zero there, even in the
presence of hyperfine structure splitting (cf. Figure \ref{fig:naw2}). 

The dashed and dotted curves in Figure \ref{fig:namodel} are models based on Equation
(\ref{eq:namodel}). The dotted curve represents standard scattering
theory, while the dashed curve is based on the solid line of Figure
\ref{fig:p21p22} for $P_{21}/P_{11}$ that was obtained with the new theory. 
To make the model parameters dimensionless, we normalize $P_{21}$ and
$P_{11}$ in terms of the value of $P_{11}$ at 5893\,\AA. The 
parameter values $a_c=17$ and $b_c=0.0035$ have been used for both the
dashed and dotted curves, but a somewhat different global scaling
factor $s$ had to be chosen for the two curves: 0.84 for the dashed
and 0.94 for the dotted one. The criterion for this choice was to make
the dashed and dotted curves coincide across the D$_2$ line and in the
far red wing of the D$_1$ line, at the same time as achieving an
excellent fit to both the blue and red wings of the D$_2$ line and to
the red wing of the D$_1$ line. It is indeed remarkable that the
simple model of Equation (\ref{eq:namodel}) allows such a close fit. 

While the D$_2$ fits of the dashed and dotted curves are identical, we
notice in Figure \ref{fig:namodel} that the dashed curve fits the
observations significantly better in the near D$_1$ wings, both on the
blue side (where the polarization is negative) and the red side of the
D$_1$ resonance. While the dotted curve of the standard scattering theory
is anti-symmetric around the D$_1$ resonance, the new 
interference terms raise the D$_1$ polarization in the positive
direction on both sides of the line, as already
indicated by the behavior of $P_{21}/P_{11}$ that we saw in Figure
\ref{fig:p21p22}. 

The simple LSA model of Equation (\ref{eq:namodel}) cannot 
be expected to be any good in the Doppler cores and their immediate
surroundings of strong resonance lines, because these optically very
thick regions are governed by intricate radiative-transfer effects,
with dramatic height and wavelength variations of the anisotropy of the radiation field, and
with profile-shaping effects of partial frequency redistribution (PRD). Thus
it is well known that the $Q/I$ profiles of strong resonance lines are
characterized by a narrow central peak in the Doppler core, surrounded
by sharp polarization minima and broader wing maxima
\citep{stenflo-reessaliba82,stenflo-holzreuteretal05}, exactly of the 
kind that we see not only in Figure \ref{fig:namodel} for the observed D$_2$
profile, but also for example for the Ca\,{\sc i} 4227\,\AA\ line
\citep{stenflo-s74,stenflo-biandaetal99} and for the Ca\,{\sc ii} K
line at 
3933\,\AA\ (Stenflo 2003). In the case of the Na D$_2$ line it was explicitly demonstrated by
\citet{stenflo-flurietal03} that the $Q/I$ triplet-type
structure with the narrow core peak could indeed be explained as the
result of effects in polarized radiative transfer with PRD. 

While the D$_1$ core peak with its surrounding sharp minima looks 
qualitatively similar to the corresponding D$_2$ line core structure,
it has previously not been possible to explain the centered D$_1$ core
peak in terms of PRD effects in polarized radiative transfer. The
reason is that with standard scattering theory one gets an
antisymmetric shape of the $P_{21}/P_{11}$ curve around the D$_1$
resonance. This has the consequence that the symmetric frequency redistribution mixes the
positive contributions from the red side of the line with the equally
large negative contributions from the blue side of the line, which leads to
cancellation of the PRD effects. The situation changes radically with
the introduction of the new interference effects, because they break
the antisymmetry by adding a contribution that is positive 
on both sides of the line. There will be no cancellation
effects when frequency redistribution acts on this symmetric component. 

A good way to understand how this works is to treat the sodium phase
matrix as the sum of two separate phase matrices: one representing 
standard theory, the other governed exclusively by the new, previously
overlooked, contributions. The two phase matrices will appear in the 
polarized radiative transfer equation as two separate scattering 
source functions. While the source function that originates from standard
scattering theory will not generate any PRD polarization peak because
of the cancellation effects of the anti-symmetric profile, the source function
that represents the new contributions can serve as a source for a positive and
symmetric polarization peak in the Doppler line core. 

\begin{figure}[t]
\resizebox{\hsize}{!}{\includegraphics{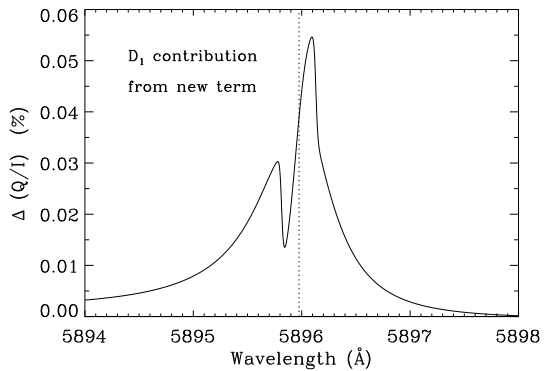}}
\caption{Illustration of the surplus polarization that is generated by
  the scattering contributions from the new term on the right-hand
  side of Equation (\ref{eq:sscript}). It represents the difference between
  the dashed and dotted model curves of Figure \ref{fig:namodel} and
  demonstrates that the new $Q/I$ contributions are positive and
  nearly symmetric around the Na D$_1$ resonance. 
}\label{fig:nadiff}
\end{figure}

The difference between the new and the old
(standard) scattering theory is only significant
around the D$_1$ line and vanishes elsewhere. Figure
\ref{fig:nadiff}, which represents the difference between the dashed
and dotted curves in Figure \ref{fig:namodel}, indicates what the main
observable consequences of the new interference terms are. Since the LSA
model cannot be expected to be very useful in the Doppler core, one
should not assign any particular significance to the odd-looking sharp
dip of the difference profile. It is an artefact of our model
simplifications and should not distract from the demonstration that the new
interferences contribute to a positive $Q/I$ polarization peak that is nearly
symmetric around the D$_1$ resonance. 

This leads to the plausible conjecture that if full polarized
radiative transfer with PRD were to be carried out for D$_1$ when the new
interferences are included, then a triplet-type
polarization structure of the observed kind (as shown by the solid
curve in Figure \ref{fig:namodel}) may emerge, because the D$_2$ line
is structured this way, and we expect that the D$_1$ and D$_2$ lines
are formed in similar ways. Only explicit
radiative-transfer modeling will answer the question whether this
conjecture is correct or not.

\section{Application to laboratory scattering in the K D$_1$
  line}\label{sec:kd1}
Observational benchmarks for testing new theories require 
laboratory experiments, where we can explore the physics under
controlled conditions. In contrast, in the solar laboratory 
the parameters are chosen by the Sun and are not directly known to the
observer. The only published laboratory 
data that are relevant to the physics of the present paper were
obtained a decade ago, based on an exploration of polarized scattering in the D$_1$ and
D$_2$ lines of potassium. The experiment has been
described in detail by
\citet{stenflo-thalmannspw4,stenflo-thalmannspw5}. It was found
that the K\,{\sc i} 
D$_1$ line has a considerable polarization structure, which proves that it is
not at all a ``null line'' as it has been predicted to be according to standard
scattering theory. This observation has led to the conclusion 
that scattering theory has to be extended to
include previously overlooked interferences, which occur because of
fine or hyperfine structure splitting of the ground state into separate substates 
\citep{stenflo-s09,stenflo-s15costarica,stenflo-s15apj}. 

Potassium rather than sodium was chosen for the experiment, because convenient solid-state
tunable lasers are available for the K\,{\sc i} D$_2$ 7665\,\AA\ and the
K\,{\sc i} D$_1$ 7699\,\AA\ wavelengths. While the experiment had many
degrees of freedom we will limit the discussion here to $90^\circ$
scattering of linearly polarized light. The Stokes
coordinate system is chosen such that the positive Stokes $Q$
direction is defined to be
perpendicular to the scattering plane. We focus our attention on two cases that
will give us the profiles of the crucial $P_{21}$ and $P_{22}$ phase
matrix elements: Scattering with the incident radiation 100\,\%\ linearly
polarized perpendicular to the scattering plane ($I+Q$ radiation) and 100\,\%\ linearly
polarized parallel to the scattering plane ($I-Q$ radiation). From
half the sum and half the difference between these two cases the
$P_{21}$ and $P_{22}$ matrix elements can be derived. 

The monochromatic incident frequency $\omega^\prime$ is scanned across
the range where the absorption resonances of either D$_1$ or D$_2$
occur, which is done by tuning the laser (with separate laser heads for D$_1$
and D$_2$). The full Stokes vector of the scattered beam can be
measured, but there is no spectral selection done in the output
arm. The detector system effectively performs an integration over
all the scattered frequencies $\omega$. 

The working temperature of the vapor cell is approximately $100^\circ$
C, which gives a thermal Doppler width of 10.2\,m\AA\  for
potassium. The cell is filled with an argon buffer gas that prevents 
diffusion of the heated potassium atoms to the cooler cell windows,
where they could cause opaque deposits. The collisions of the
potassium atoms with the buffer gas however lead to both collisional
broadening and depolarization effects, which greatly complicates the
interpretation of the measurements. We therefore need to
make use of a theory for partial frequency redistribution (PRD) with
the collisional effects entering via different branching ratios as
well as in the form of line broadening and collisional depolarization
or decoherence. The problem is that such a theory only exists for the
2-level case with standard scattering. Our only option here is to make
a phenomenological extension of this PRD theory to the multi-level
case, and in particular define in a parametrized and 
heuristic way how collisions affect the new interference terms. The
details of this is 
dealt with in Appendix \ref{sec:branching}. While the presence of
collisions seriously 
complicates the quantitative modeling of the 
polarization effects, the qualitative aspects, sign behavior, and
orders of magnitude are not affected. We recall that standard
scattering theory predicts null results for the cases that we
will consider, regardless of the PRD theory used. It can therefore
never be brought into any agreement with the observations.

\subsection{Mueller matrix of the experiment}\label{sec:optdepth}
The temperature of the vapor cell determines the number density of
potassium atoms and is chosen as a compromise: high to
give sufficient scattering probability, but not too high to avoid
multiple scattering effects. As the optical depth of the cell scales
with the number density, the optimum compromise setting results in an
effective optical depth that is smaller, but not much smaller, than
unity. If we define $\tau$ as the optical depth at line center, where
the intensity profile 
$\Phi_I$ has a maximum, then as explained in \citet{stenflo-s15apj} we find from model
fitting that the measurements were done with $\tau=0.25$ for D$_1$
(and therefore with $\tau=0.50$ for D$_2$, due to the twice larger
oscillator strength of the D$_2$ line). 
This value of $\tau$ was needed to explain the observed field strength
sensitivity of the D$_1$ $Q$ polarization in the case of $+Q\to Q$
scattering (incident radiation linearly polarized perpendicular to the
scattering plane), because for vanishing $\tau$ the scattered $Q$ has almost
no field dependence. 

The optical depth effects are so small that they may be ignored for
the non-magnetic cases of scattering of linearly polarized light that
we will consider here, but we include them here
for completeness. They have been accounted for in the model
computations. Extending the approach of \citet{stenflo-s15apj}, we
express the 
effective Mueller matrix of the scattering experiment as 
\begin{equation}
\vect{M}_{\rm eff}(\omega,\omega^\prime)=\vect{M}_{\rm arm}(\omega)\,\,\vect{R}(\omega,\omega^\prime)\,\,\vect{M}_{\rm arm}(\omega^\prime)\,,
\label{eq:muellereff}\end{equation}
where $\vect{R}(\omega,\omega^\prime)$ is the redistribution matrix in
the laboratory frame, and 
\begin{equation}
\vect{M}_{\rm arm}=e^{-\vect{\Phi}\,\tau}
\label{eq:muellerarm}\end{equation}
is the Mueller matrix for the input or output cell arm. $\vect{\Phi}$
is the area-normalized standard Mueller absorption matrix
(cf. Appendix \ref{sec:incoh}). The exponentiation
of a matrix is as usual defined by its Taylor expansion. 

In the experimental
setup the input and output are treated in fundamentally different
ways. The Mueller matrix $\vect{M}_{\rm obs}$ that is actually
measured is the one that is obtained after the
detector system has integrated over all the scattered frequencies: 
\begin{equation}
\vect{M}_{\rm obs}(\omega^\prime)=\int\vect{M}_{\rm
  eff}(\omega,\omega^\prime)\,\,{\rm d}\omega\,.
\label{eq:muellerobs}\end{equation}
This gives us all the Mueller matrix elements as functions of the
incident frequency that is tuned by the laser.

\subsection{Model Fitting}\label{sec:modelfit}
The phenomenologically extended PRD theory that we use for dealing with
the collisional effects to model the laboratory experiment is defined in
Appendix \ref{sec:branching}. The free parameters of the model are the
elastic collision rate $\Gamma_E$, the rate of destruction of the
2$K$-multipole $D^{(K)}$, and the optical depth $\tau$. In addition we
have for the new interference terms introduced a scaling factor $k_g$, which physically
represents an enhanced vulnerability to
collisional destruction of the polarization. 

The ratio $D^{(K)}/\Gamma_E$ is 0.5 in classical scattering theory
\citep{stenflo-bs99}. Although it can assume other values in quantum physics, the
values do not tend to deviate that much from 0.5. Therefore we
eliminate one free parameter by fixing the ratio $D^{(K)}/\Gamma_E
=0.5$. 

As mentioned in the previous subsection, a non-zero value of $\tau$ is
needed to explain the observed magnetic-field sensitivity of linearly
polarized scattering. Modeling in \citet{stenflo-s15apj}  fixed its value to $\tau=0.25$
for K D$_1$ (while it is 0.5 for K D$_2$). Apart from contributing
slightly to
the line broadening it has no other significant effect on the
non-magnetic scattering that we will consider here. 

We are therefore left with only two remaining free parameters,
$\Gamma_E$ and $k_g$. While $k_g$ is significant for D$_1$ scattering,
its effect on D$_2$ is completely negligible, since the D$_2$
polarization is governed almost exclusively by standard scattering
theory. As the D$_2$ polarization amplitude
decreases strongly and monotonically with increasing $\Gamma_E$ (due to the collisional
depolarization), $\Gamma_E$ gets uniquely determined by fitting the
observed D$_2$ polarization amplitude. 

\begin{figure}[t]
\resizebox{\hsize}{!}{\includegraphics{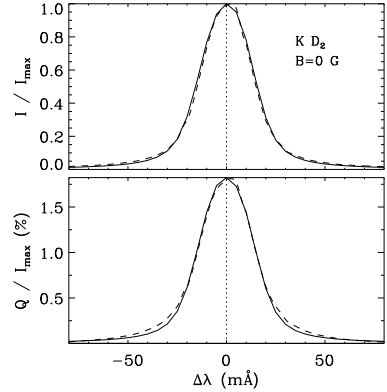}}
\caption{Theoretical fit (solid curves) of the Stokes $I$ (upper
  panel) and $Q$ (lower panel) profiles that have been observed in the laboratory (dashed
  curves) for
  non-magnetic $90^\circ$ scattering at potassium gas with the incident
  radiation 100\,\%\ linearly polarized perpendicular to the scattering
  plane. The scale of the D$_2$ polarization is governed 
  by the elastic collision rate $\Gamma_E$, which for this fit is 
  77 times the natural radiative $\Gamma$. With this damping and the
  known thermal Doppler broadening, we automatically achieve a perfect
  fit of the line width, without the need for any additional
  broadening mechanism. 
}\label{fig:kd2}
\end{figure}

Figure \ref{fig:kd2} shows the nearly perfect fit between model (solid
curves) and observations (dashed curves) for the case of non-magnetic
D$_2$ scattering, when the incident radiation is linearly polarized
perpendicular to the scattering plane. $\Gamma_E
=77\,\Gamma$ for this fit, where $\Gamma$ is the radiative damping constant. Note
that although the model only aims at reproducing the observed
amplitude with the single free parameter $\Gamma_E$, the profile
widths and shapes get automatically reproduced as well, without any
additional assumptions or parameters. The width is determined by the
thermal Doppler broadening, which is known and therefore fixed, the
damping width that is dominated by $\Gamma_E$, and to some 
extent by $\tau$ (which was fixed by the field-sensitivity criterion
for D$_1$).  With a significantly smaller value for $\Gamma_E$ we
would have had to assume some additional, ad hoc broadening mechanism
to achieve agreement between the profile widths of the model and the
observations. With a significantly larger $\Gamma_E$ no fit of the
profile width would be possible at all. This demonstrates the full
consistency of the parameter fit. 

\begin{figure}[!t]
\resizebox{\hsize}{!}{\includegraphics{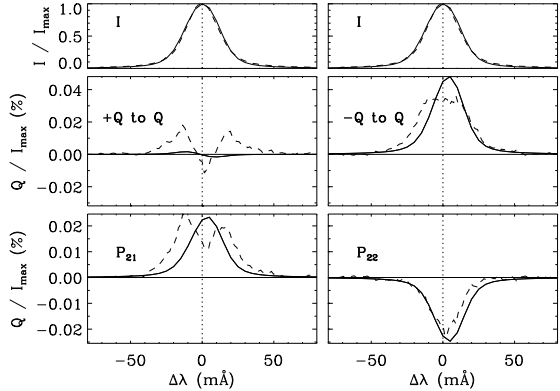}}
\caption{Theoretical model fits (solid curves) of laboratory K D$_1$
  scattering data for the same model parameters as used for the K D$_2$ fit
  in Figure \ref{fig:kd2}, while choosing $k_g=0.4$ for the special
  depolarization parameter of the D$_1$ model. The experimental data,
  from  \citet{stenflo-thalmannspw4},  are the same as illustrated in Figure
  \ref{fig:SPW4_6}, except that we have here added a correction for an assumed
  $I\to Q$ cross talk of $2.0\times 10^{-4}$. This
  correction has however no effect on the $P_{22}$ panel at the bottom
  right. The main signature of the new interference effects is the
  opposite signs of the $P_{21}$ and 
  $P_{22}$ matrix elements, while standard theory predicts null results. 
}\label{fig:spw4}
\end{figure}

We can now do the D$_1$ model fitting with only one free 
parameter, $k_g$. The observations that we want to reproduce are the
ones that have been illustrated in Figure \ref{fig:SPW4_6}, which was adapted from 
\citet{stenflo-thalmannspw4} and represents non-magnetic 
scattering of linear polarization in the K D$_1$ line. Here we only
use the results obtained with full laser power, reproduced as the
dashed curves in Figure \ref{fig:spw4}, because they are
the same as the ones obtained with ten times less power. We have also
added an instrumental $I\to Q$ cross talk correction to the
experimental data, as explained below. The
theoretical profiles are given by the solid curves. 

The panel labeled ``$+Q$ to
$Q$'' represents the measurements when the input radiation is 100\,\%\
linearly polarized perpendicular to the scattering plane, the
panel labeled ``$-Q$ to 
$Q$'' when the input radiation is 100\,\%\
linearly polarized parallel to the scattering plane. These two panels
therefore represent the combinations $P_{21}\pm P_{22}$, respectively,
of the phase matrix elements, normalized to $P_{11,{\rm max}}$. Due to the large collisional
depolarization $\vert P_{12}\vert$ can be neglected in comparison with
$P_{11}$, which implies that Stokes $I$ is represented by $P_{11}$
alone. The observational curves in the bottom panels for $P_{21}$ and 
$P_{22}$ are then obtained by taking half the sum and half the 
difference of the middle panels. 

Like for D$_2$ the Stokes $I$ profile width depends significantly on
the collisional damping width $\Gamma_E$. The nearly perfect model fit
to the D$_1$ Stokes $I$ profile in the top panels validates the
self-consistency of the $\Gamma_E$ parameter as determined from the
D$_2$ observations. 

In contrast to the observations, the theoretical model calculations give the phase matrix
elements of the bottom panels directly, from which the $\pm Q$ to $Q$ versions of the
middle panels are then derived. Comparison between the theoretical and
observational curves shows qualitative agreement as concerns the
relative signs of the phase matrix elements and the order of
magnitude of their amplitudes, but the agreement is significantly better for the
$P_{22}$ diagram, which is unaffected by $I\to Q$ instrumental cross
talk, than for the other three diagrams of the middle and bottom
panels of Figure \ref{fig:SPW4_6}, when they are uncorrected for the
cross talk. This leads us to suspect that
some of the discrepancies have to do with such cross talk. 

Because Stokes $I$ is so much larger than the other Stokes parameters,
the instrumental polarization is dominated by the $I\to Q$ cross talk,
which may arise from stresses in the exit window of the vapor cell. It can
be accounted for in the model by adding a fraction $f$ of the $I$
profile of the top panels to the measured $Q$ of the middle
panels. This $I$ fraction then also gets added to the bottom left
panel for $P_{21}$, because it is obtained as the average of the two
middle panels, but it gets subtracted away in the bottom right panel
for $P_{22}$, which is formed from half the difference between the
middle panels. We find that the application of a cross talk fraction $f=2.0\times
10^{-4}$ significantly improves the fit. This (quite small) cross talk correction
has therefore been applied to the experimental data in Figure
\ref{fig:spw4}. 

Besides the cross talk correction we find that the model amplitudes
would be too high by a factor between 2 and 3 unless we make use of
the free scaling parameter $k_g$. A
value of $k_g=0.4$ has therefore been used in Figure
\ref{fig:spw4} to optimize the fit. It may be interpreted
as an enhanced vulnerability to collisional depolarization as compared
with the depolarization that is expected from the standard
branching ratios in PRD theory. 

The main property that is well expressed by our model is the observed
symmetry breaking between the $P_{21}$ and $P_{22}$ phase matrix
elements. According to standard scattering theory they should both be
zero. In contrast, the observations show $P_{22}$ to be negative,
$P_{21}$ to be positive. 

Nevertheless there still remain important modeling problems. The profile
shapes are not fully reproduced even after introducing the correction
for $I\to Q$ cross talk, and the $k_g$ scaling represents an extra
free parameter that cannot be derived from the theory. While we still
do not have a full understanding of D$_1$ 
scattering physics, much of the present modeling complications have to do with
the circumstance that the scattering takes place  in a
collision-dominated regime. A heuristic 
extension of phenomenological polarized frequency redistribution
theory cannot be expected to properly describe the detailed
redistribution physics in this regime in the presence of new,
previously unexplored, interference effects. We need a laboratory
experiment for the collision-free 
regime to eliminate such irrelevant complications, so that we may more
directly address the fundamental issues
concerning the quantum scattering of light. 

One objection that has been raised about our experimental D$_1$
results is that there could be collisional transfer of atomic
polarization from the upper state of D$_1$ to that of D$_2$, which is then 
radiated away in the D$_2$ line. Since we do not have spectral
discrimination in the output arm of the experiment, we cannot
distinguish such D$_2$ photons from those that are scattered within
the D$_1$ transition. However, this objection is unfounded, because
the velocity distribution of the buffer gas colliders is thermal and
therefore isotropic. Without any preferred spatial orientation the
collisions are incapable of transferring any polarization, even if
significant D$_1$ to D$_2$ collisional transfer would take place
(which is also highly questionable). Although we therefore believe
that collisions with the buffer gas can be completely ruled out as a
possible source of polarization, this kind of debates would be
unnecessary for an experiment in the collision-free domain.

\section{Outlook}\label{sec:outlook}
Scattering of light is a cornerstone process in physics that needs to
be deeply understood in all of its aspects. Although laboratory
experiments on polarized scattering was a hot topic in the early days
of quantum mechanics, because they allowed explicit  demonstration of the
fundamental concept of coherent superposition of atomic states in
various situations, the
scientific community turned to other topics around 1935, apparently
because it was felt that the experimental possibilities had been
exhausted with the technology that was available at that time. This
technology was incredibly crude by today's standards. Using tunable
lasers as light sources, electro-optical polarization modulators, and
photoelectric detectors we can now do orders of magnitude better and
explore parameter domains that were entirely inaccessible at that
time. 

The motivation to return
to this topic after it had for many decades been 
considered to be sufficiently understood, came from observations of polarization
anomalies in the Sun's spectrum. For the resolution of the D$_1$
enigma one needed a 
laboratory experiment under controlled conditions. Such an experiment 
was carried out a decade ago for 
polarized scattering at potassium vapor. Although it unequivocally revealed that
available scattering theory is indeed incomplete, the general
reaction of the community was one of unsubstantiated disbelief. The
feeling was that the theory for the scattering of light has been used
for so many decades that it cannot be wrong and does not need to be
tested. Therefore there must be something wrong with the experiment
(for instance that there must be some special effects when using laser light,
collisions may transfer polarization to the D$_2$ line, or there are
some unidentified instrumental problems). Instead of trying 
to examine if any of these various objections could have some merit,
the reaction to the ``inconvenient evidence'' was rather to look away
and continue with ``business as usual''.  

This behavior brings to mind the wise words of the eminent
Austrian-British science philosopher Karl Popper in a quote from 1957:
``If we are uncritical we 
shall always find what we want: we shall look for, and find 
confirmations, and we shall look away from, and not see, whatever
might be dangerous to our pet theories.'' For Popper the most
important element for progress in experimental science is the
falsification of theories, and not their verification. In the context
of the D$_1$ problem it is the falsification of standard scattering
theory that is the by far most important aspect, while the degree of
success of the new modeling attempts is secondary. After the
falsification aspect has now been taken care of, a next generation of
laboratory experiments will be needed to guide the development of an 
extended scattering theory that can satisfy all observational
constraints in parameter domains where it has not been experimentally
tested before.

\subsection{Next steps}\label{sec:next}
In the solar case the application that we have done in the present
paper to compare the
extended scattering theory with
the observations is incomplete, because proper modeling of the
solar line profiles requires radiative transfer with partial frequency
redistribution. Fortunately the sodium D$_1$ line core is formed at
heights in the solar atmosphere, where collisional effects are rather
unimportant, which justifies the collision-free application of the
theory. As collisional branching ratios are then not needed, the
redistribution matrix $\vect{R}$ in Equation (\ref{eq:riiriii}) is
given exclusively by matrix $\vect{R}_{\rm  II}$, which greatly
simplifies the radiative transfer problem. We may expect that
redistribution acting on the
nearly symmetric profile of Figure \ref{fig:nadiff} for the surplus D$_1$
polarization, which is generated by the new interference effects,  will lead to a final profile
shape that is qualitatively similar to that of the D$_2$ profile, as
needed for agreement with the observed D$_1$ profile shape. This
conjecture needs to be tested. Still the solar laboratory is 
inferior to terrestrial laboratories for the purpose of unambiguous tests of
theories, because we have no control over the physical parameters on the Sun. 

The laboratory experiment done so far has for convenience used a
potassium vapor cell with an argon buffer gas, which led to undesired
high collision rates causing line broadening and 
depolarization. It then became unavoidable to do modeling in the
collision-dominated domain with heuristically
extended PRD physics. For the next
generation of scattering experiments the first priority should be
to avoid such complications in order to do ``clean'' tests of the
various 
aspects of scattering physics. The experiment should therefore be done in the
collision-free regime without the use of any buffer gas. Once 
this regime has been sufficiently explored, including the effects of imposed
external magnetic fields, one would have a foundation that could 
later be extended by exploring collisional physics
and PRD in the laboratory. The various
applications of scattering theory, e.g. in astrophysics, need to be
based on a solid foundation that should always be validated in the 
laboratory, as we have learnt from the example of the D$_1$ enigma. 

The next generation of experiments with polarized scattering are being
considered in collaboration with INLN (Institut Non-Lineaire de Nice)
in France, where they have facilities and experience for doing various
kinds of atomic-physics experiments with rubidium vapor without the
use of any buffer gas \citep[cf.][]{stenflo-baudouin14,stenflo-guerinetal16}. It is
possible to do the experiment with the 
natural mixture of the two main Rb isotopes 85 and 87 or with each of
the isotopes separately. External magnetic fields of various strengths
and orientations may be imposed, and the rubidium may be supercooled
to microkelvin temperatures if desired (to eliminate all Doppler
broadening). This would open the door to a vast parameter space to
achieve in-depth clarification of the various aspects of
scattering theory. 

\begin{figure}[t]
\resizebox{\hsize}{!}{\includegraphics{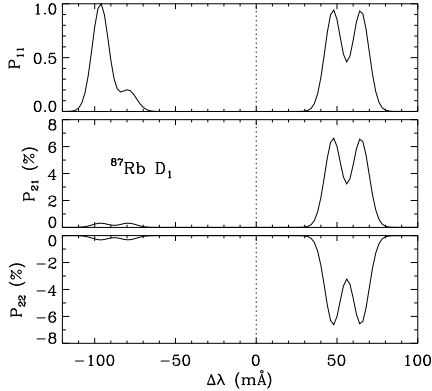}}
\caption{Predicted phase matrix elements $P_{11}$, $P_{21}$, and
  $P_{22}$, given in units of $P_{11,\,{\rm max}}$, for $90^\circ$ D$_1$ 
  scattering at rubidium isotope 87 (which has
  nuclear spin 3/2). Note the perfect anti-symmetry between
  $P_{21}$ and $P_{22}$. The reference wavelength for the relative
  wavelength scale is 7947.6712\,\AA. 
}\label{fig:rb87}
\end{figure}

\begin{figure}[t]
\resizebox{\hsize}{!}{\includegraphics{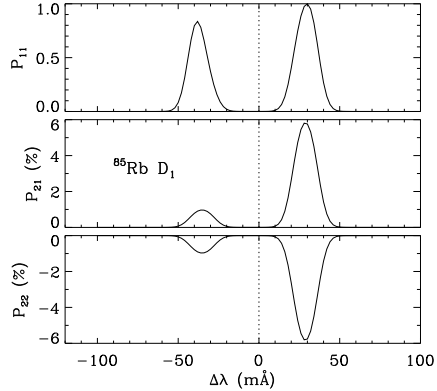}}
\caption{Same as Figure \ref{fig:rb87} but for D$_1$ scattering at rubidium
isotope 85 (which has nuclear spin 5/2).The reference wavelength for
the relative wavelength scale is the same as in Figure \ref{fig:rb87}.}\label{fig:rb85}
\end{figure}

In anticipation of the future laboratory experiments we apply the
present version of our extended scattering theory to make a
definite prediction in Figures \ref{fig:rb87} and \ref{fig:rb85} of what
we may expect to find in the case of 
non-magnetic, collision-free $90^\circ$ scattering at rubidium
vapor. The prediction is illustrated for the phase matrix elements
$P_{11}$ (normalized intensity) and $P_{21}$ and $P_{22}$, which govern
the linear polarization and are the matrix elements that are relevant to the
solar D$_1$ enigma and the K D$_1$ experimental results that were
illustrated in Figure \ref{fig:spw4}. As before, the Stokes coordinate
system is defined such that the positive Stokes $Q$ direction is
perpendicular to the scattering plane. 

The natural isotope composition of rubidium is 72.2\,\%\ of isotope 85
with nuclear spin 5/2,  and 27.8\,\%\ of isotope 87 with nuclear spin
3/2. The isotopes may be purified to be scattered at separately, but
if the experiment is done for the natural isotope mixture, then the
predictions of Figures \ref{fig:rb87} and \ref{fig:rb85} should be
weighted together in proportion to the corresponding relative
abundances after the relative isotope shift has been accounted for. The isotope shift
of +1.46\,m\AA\ of isotope 85 relative to isotope 87 
has already been applied to Figure \ref{fig:rb85} (although
it is too small to be visible in the plot), so that both relative
wavelength scales in fact refer to the same reference wavelength, 7947.6712\,\AA.  

Rb isotope 87 has the same nuclear spin as sodium and potassium, but
the hyperfine structure splitting (HFS) is approximately 16 times
larger for Rb than for potassium. Therefore the Rb HFS
components are so widely separated that there is hardly any overlap
between them.  In the case of the K lab experiment we had in addition
collisional line broadening by a factor of 77, which greatly
contributed to the overlap of the profiles from the
different HFS components. Such overlap does not occur in the
collision-free Rb case. 

We notice in both of Figures \ref{fig:rb87} and \ref{fig:rb85} the perfect
anti-symmetry between the $P_{21}$ and 
$P_{22}$ matrix elements. This predicted anti-symmetry is a special
signature of the new interference terms in our present version of the extended
theory. It should be stressed, however, that any observed non-zero
features in the spectra of $P_{21}$ and $P_{22}$ would falsify
standard scattering theory, which predicts both of them to be zero for
all wavelengths. The laboratory experiments are needed to guide us in
the difficult process of formulating a theory that is able to satisfy
all observational constraints.

\begin{acknowledgements}
For the modeling of the scattering polarization I have greatly
profited from an IDL code provided to me by Svetlana Berdyugina and
Dominique Fluri, with which one can conveniently calculate the atomic
level structure and transition 
amplitudes in the Paschen-Back regime for the sodium and potassium
D$_1$ -- D$_2$ systems, and from an extension of this code by K. Sowmya to the
rubidium isotopes 85 and 87. 
\end{acknowledgements}

\appendix

\section{Rabi frequency for the laboratory experiment on potassium
  scattering}\label{sec:rabi}
The Rabi frequency $\omega_R$ for a given transition between $m$
states $i$ and $j$ is 
\begin{equation}
\omega_R ={\vect{d}_{ij}\cdot\vect{E}_0\over\hbar}\,,\label{eq:omrabi}
\end{equation}
 where $\vect{d}_{ij}$ is the electric dipole moment of the
transition, and $\vect{E}_0$ is the amplitude of the electric vector
of the external radiation field. $\vect{d}_{ij}\cdot\vect{E}_0$
represents the interaction energy between the atom and the radiation
field. 

Let $\cal{P}_{\rm laser}$ be the power of the laser beam, which is expanded to
fill the effective cross section $\sigma_{\rm cell}$ of the vapor cell
where the scattering takes place. Then 
\begin{equation}
{{\cal{P}_{\rm laser}}\over c\,\,\sigma_{\rm cell}}=\,\epsilon_0
\vert E_{0q}\vert^2\,,\label{eq:radevect} 
\end{equation}
where the left-hand side represents the radiative energy density of
the laser beam, and $E_{0q}$ is the polarization component of the
100\,\%\ polarized beam, which induces the $ij$ atomic transition. The
corresponding component of the dipole moment that is radiatively
excited can be expressed in terms of the Einstein $A_{ji}$
coefficient \citep[cf.][]{stenflo-loudon}: 
\begin{equation}
d_q=(3\pi\epsilon_0 \,c^3 \,\hbar\, A_{ji}/\omega_0^3\,)^{1/2}\,.\label{eq:dqaji} 
\end{equation}
Combining Equations (\ref{eq:omrabi}) -- (\ref{eq:dqaji}) we find 
\begin{equation}
\omega_R =\Bigr(\,{3\over 4\pi^2}\,\,{{\cal{P}_{\rm laser}}\,\,\lambda^3\over c\,
  \hbar\,\sigma_{\rm cell}}\,\,A_{ji}\,\Bigl)^{\!1/2}\,.\label{eq:omrfinal}  
\end{equation}
With the maximum laser power 15\,mW for $\cal{P}_{\rm laser}$, 7698.97\,\AA\ for
  the K D$_1$ resonance, 1\,cm$^2$ for the cell cross section
  $\sigma_{\rm cell}$, and $3.82\times 10^7$\,s$^{-1}$ for the
$A_{ji}$ coefficient, we find $\omega_R\approx 7.9\times
10^7$\,s$^{-1}$, approximately 2.1 times larger than $A_{ji}$. 

Converted to wavelength units, the radiative $\Gamma$, taken as
$A_{ji}$, is 0.12\,m\AA, while the Rabi frequency corresponds to
0.25\,m\AA. Model fitting of the K D$_2$ line gives an elastic
collision rate $\Gamma_E \approx 77\Gamma$. The total damping
width $\Gamma +\Gamma_E$ then corresponds to 9.4\,m\AA, which is comparable
to the thermal Doppler width (10.3\,m\AA) and 38 times larger than the
corresponding Rabi frequency. 

To rule out a possible dependence of the scattering polarization on
the Rabi frequency we have compared results obtained with full laser
power 
and with the power reduced by a factor of ten. As documented in Figure
\ref{fig:SPW4_6} one cannot discern any significant difference between
the corresponding polarization profiles, which verifies that the
relatively high radiation energy density of the laser beam does not
affect the polarization phenomena that we are exploring. 

\section{Expression for the transition matrix
  elements}\label{sec:matelement}
The matrix elements that define the atomic transitions in the
Kramers-Heisenberg formulation can be expressed algebraically with the
help of the Wigner-Eckart theorem and expansions of the reduced matrix
elements $\langle F\,\vert\vert \,e\,{\vect{\hat r}}\,\vert\vert F^\prime
\rangle$ and $\langle J\,\vert\vert \,e\,{\vect{\hat r}}\,\vert\vert J^\prime
\rangle$, using well-known formulae given for instance in
\citet{stenflo-book94} and \citet{stenflo-steckna}. The resulting 
expression for the non-magnetic case is 
\begin{eqnarray}
&\langle F\, m\vert \,e\,{\hat r}_q\vert F^\prime m^\prime
\rangle\,=\,\langle L\,\vert\vert \,e\,{\vect{\hat r}}\,\vert\vert L^\prime
\rangle \,(-1)^{L+S+I+1}\,\sqrt{2L+1}\nonumber\\ &(-1)^{J+J^\prime
  +2F^\prime+m}\,\sqrt{(2J+1)(2J^\prime +1) (2F+1)(2F^\prime
  +1)}\nonumber\\ &\begin{Bmatrix} L & \,\,L^\prime & \,\,1 \\ J^\prime & J &
  \,S \end{Bmatrix}\begin{Bmatrix} \!J & \,J^\prime & \,\,1 \\ F^\prime & F &
  \,I \end{Bmatrix}\begin{pmatrix} F^\prime & \,1 & \,\,\,F \\ m^\prime & q &
  -m \end{pmatrix}\,.
\label{eq:matel} 
\end{eqnarray}
Since $L$, $L^\prime$, $S$, and $I$ are the same for all the members
of a given supermultiplet, the three factors in the first row of the
above equation are of no consequence for the polarization, because they
divide out when forming the phase matrix (which is usually normalized
to the maximum value of $P_{11}$). However, it is essential that all
the remaining factors are included correctly. If they are not, the
Principle of Spectroscopic Stability (PSS) will not be satisfied for
all supermultiplets (all combinations of electronic and nuclear spins
$S$ and $I$). PSS is therefore a powerful tool to verify the
correctness of the expressions and to reveal possible bugs in the computer
algorithms. In the present paper we have applied it to verify that it
is obeyed for all $L=0\to 1\to 0$ scattering transitions for any
combination of $S$ (which defines the set of $J$ quantum numbers) and
$I$ (which defines the set of $F$ quantum numbers).

\section{PRD and Collisional Branching
  Ratios}\label{sec:branching}
Collisions are generally regarded as discrete events in the impact
approximation, but a consistent quantum-mechanical treatment is
extremely complex with implicit approximations. Moreover, the theory has 
not been sufficiently tested in the laboratory. Most of the
expressions used today in astrophysical contexts are based on the
early work of \citet{stenflo-omont72,stenflo-omont73}, developed into the currently
used form in particular by
\citet{stenflo-bommier97a,stenflo-bommier97b}. Here we will summarize
the way in which collisions are currently believed to affect the
scattering matrix. 
 
Collisions change the scattering matrix in two main
ways: (1) They govern the relation between the incident and scattered
frequencies $\omega^\prime$ and $\omega$ (frequency redistribution),
and (2) they change the polarization of the scattered radiation. The
effects of frequency redistribution and collisional depolarization are
coupled to each other, but a PRD theory that describes this exists
only for the 2-level standard scattering case. We therefore need to
make a phenomenological extension of PRD for standard scattering
theory to the multi-level case, and in addition introduce some
reasonable way to do the calculations for the new scattering terms. In
standard PRD theory the net effect of collisions is that each
term in Equation (\ref{eq:wmatnew}) for the coherency matrix (or
correspondingly each term of the Mueller matrix) can be described as a
sum of two main terms, one that represents frequency coherence (which we
refer to as FC and which implies that $\omega^\prime=\omega$ in the
atomic rest frame), while the
other term represents complete frequency redistribution (which we
refer to as CRD). The relative contribution  of each is governed by a
branching ratio, which we denote $A$ for FC and $B$ for CRD. The
weighted combination of FC and CRD describes the full effect of the
collisions and is referred to as PRD (partial frequency
redistribution). We also have to properly attach decoherence factors
due to interferences between the scattering amplitudes. 

In the following subsection we will outline how this standard PRD
theory can be phenomenologically extended to
multi-level scattering. For non-standard scattering with the new
interferences as defined by the second term on the right-hand side of Equation
(\ref{eq:sscript}) we need to introduce some further modifications of standard PRD
theory to allow us to include collisional effects in a way that is well defined
for modeling purposes. This heuristic extension of PRD theory is dealt with in
Subsection \ref{sec:prdnewterms}.

\subsection{Phenomenological extension of PRD for standard scattering
  theory}\label{sec:prdext} 
Let us by symbol ${\cal P}_{\rm standard}(\omega,\omega^\prime)$ denote the PRD
expression for the frequency-dependent parts of the bilinear
products in Equation (\ref{eq:wmatnew}) for a given combination
$a,a^\prime\!,b,b^\prime\!,f,f^\prime$ of initial, intermediate, and final
states in standard scattering theory. Then 
\begin{eqnarray}
\label{eq:pprd} 
&{\cal P}_{\rm standard} (\omega,\omega^\prime) =\textstyle{1\over
  2}[\Phi_{\,b-a}(\omega^\prime)+\Phi_{\,b^\prime-a}^\ast(\omega^\prime)]
\cos\beta_e\,e^{i\beta_e}\\
&\Bigl[\,A\,\delta(\omega-\omega^\prime) \,+\,B^{(K)}
\,\textstyle{1\over 2}[\Phi_{\,b-f}(\omega) 
\,+\Phi^\ast_{\,b^\prime-f}(\omega)]
\cos\alpha_e\,e^{i\alpha_e}\Bigr]\nonumber\,.  
\end{eqnarray}
Let $\Gamma$,  $\Gamma_I$, and $\Gamma_E$ be the damping constants
of radiation, inelastic collisions, and elastic collisions,
respectively. Then the profile functions $\Phi$ are given by Equation
(\ref{eq:phiba}) if we do the replacement 
\begin{equation}
\label{eq:gamc} 
\gamma=\Gamma +\Gamma_I +\Gamma_E\,. 
\end{equation}
Index $e$ of the decoherence angles $\beta_e$ and $\alpha_e$ indicate
that the decoherence originates exclusively from level splitting of 
the excited state. $\beta_e$, which relates to the absorption process, is defined
by 
\begin{equation}
\tan\beta_e={\omega_{\,b^\prime b}\over \Gamma +\Gamma_I +\Gamma_E}\,,
\label{eq:tanb}\end{equation}
while $\alpha_e$ that relates to the emission process is defined
by  
\begin{equation}
\tan\alpha_e={\omega_{\,b^\prime b}\over \Gamma +\Gamma_I +D^{(K)}}\,,
\label{eq:tana}\end{equation}
where $D^{(K)}$ represents the destruction rate of the
$2K$-multipole (the atomic polarization) when $K=1$ or 2. 

Comparison of Equations (\ref{eq:tanb}) and (\ref{eq:tana}) shows that
there is a fundamental difference between the absorption and emission
processes (in contrast to the collisionless case), because the
collisional effects  enter
differently in the two cases: as $\Gamma_E$ and as $D^{(K)}$,
respectively. This implies that the collisions manage to break the
symmetry with respect to time reversal, although it remains rather obscure
how they actually do it. 

The branching ratios are 
\begin{equation}
A ={\Gamma \over \Gamma + \Gamma_I + \Gamma_E}
\label{eq:brii} 
\end{equation}
and 
\begin{equation}
B^{(K)} =A\,\,\,{\Gamma_E -D^{(K)} \over \Gamma + \Gamma_I + D^{(K)}}\,.
\label{eq:briii} 
\end{equation}
If there were symmetry between the absorption and emission processes,
then $B^{(K)}$ would be exactly zero, and all scattering would be
frequency coherent. 

The assignment of different combinations of dipole
transitions to different values of $K$ is a technical issue that we will not go
into here but only refer to the treatment in
\citet{stenflo-sampoorna07}. Both classical and quantum collision
theory suggest that 
$D^{(K)}$ should usually be of order $\Gamma_E/2$ for both $K=1$ or 2
\citep{stenflo-bs99,stenflo-spielfiedel91,stenflo-faurobetal95}, although
the value depends on the atomic structure and it is even questionable
to what extent the parametrization of the collisional effects in terms of the
two parameters $D^{(1)}$ and $D^{(2)}$ represents a good approximation. 

From Equation (\ref{eq:briii}) we see that $B^{(K)}\approx A$ for
large collision rates (except in the unlikely special case when
$D^{(K)}\approx \Gamma_E$), while $B^{(K)}\approx 0$ when $\Gamma_E \ll \Gamma +
\Gamma_I$. 

$K=0$ represents the isotropic case without atomic or scattering
polarization. Since $D^{(0)}=0$,  and since we may safely assume that
$\Gamma_I\ll \Gamma_E$, Equation (\ref{eq:briii}) gives 
\begin{equation}
B^{(0)} \approx{\Gamma\over \Gamma + \Gamma_I }\,\,(1-A)\,.
\label{eq:bk0} 
\end{equation}
This represents the case of collisionally induced isotropic,
unpolarized scattering, when the collisions have erased all atomic
``memory'' of the excitation event, so that there is no phase or
directional relations between absorption and emission processes. It
will be clarified more in Section \ref{sec:incoh} under the name {\it incoherent
scattering}. 

In terms of the Mueller scattering matrix $\vect{M}$, $K=0$ corresponds to
the isotropic component $M_{11}$,  while all the remaining
parts of $\vect{M}$ relate to either $K=1$ or 2. 

What we usually refer to as {\it redistribution matrices} $\vect{R}$
are Mueller matrices that are formed from the coherency matrix
$\vect{W}$ after summing up the contributions from all the index
combinations of $a,a^\prime\!,b,b^\prime\!,f,f^\prime$, making use of
Equation (\ref{eq:pprd}). In general one also needs to apply standard Doppler redistribution
integrals to transform from the atomic rest frame to the observer's
coordinate system if one wants to get the correct spectral
distribution of the scattered radiation. In the special case of our
laboratory scattering experiment, however, the detector system
effectively integrates over all the scattered frequencies. Then it is
sufficient to do gaussian Doppler broadening for the incident
frequencies over which the laser tuning is done, while ignoring how
the frequencies get redistributed by the scattering process. 

As in Equation (\ref{eq:pprd}) $\vect{R}$ can be decomposed in a
frequency coherent and a complete 
redistribution part, for which the respective notations $\vect{R}_{\rm
  II}$ and $\vect{R}_{\rm III}$ have been used in the literature. Thus we may
write 
\begin{equation}
\vect{R}=A \,\vect{R}_{\rm  II} +\sum_{K=1}^2\,B^{(K)} \,\vect{R}_{\rm III}^{(K)} + \,{\Gamma\over \Gamma + \Gamma_I }\,(1-A)\,\vect{R}_{\rm ic}\,,
\label{eq:riiriii} 
\end{equation}
where we for convenience have introduced the notation $\vect{R}_{\rm
  ic}$ to represent the isotropic part $\vect{R}_{\rm III}^{(0)}$
(using index ``ic'' to indicate that this matrix represents incoherent
scattering).  

The normalization condition for $\vect{R}$ is 
\begin{equation}
\int{{\rm d}\Omega^\prime\over 4\pi}\,\int{{\rm d}\Omega\over 4\pi}\,
\int{\rm d}\omega^\prime\int{\rm d}\omega\,\,\vect{1}^T \vect{R}\,\vect{1}\,=1\,,
\label{eq:rnorm} 
\end{equation}
where
\begin{equation}
\vect{1}^T \vect{R}\,\vect{1}=R_{11}\,.
\label{eq:r11} 
\end{equation}
$R_{11}$ is the first element of matrix $\vect{R}$, while $\vect{1}$
is a 4-vector that has unity in its first position while the rest is 
zero. Upper index $T$ means transposition. The angular averaging is made
over all incident and scattered directions, the frequency integrations over all
incident and scattered frequencies.

\subsection{Heuristic extension of PRD for the new scattering
  terms}\label{sec:prdnewterms} 
In analogy with Equation (\ref{eq:pprd}) we let symbol ${\cal P}_{\rm
  new}(\omega,\omega^\prime)$ denote the PRD expression for the
frequency-dependent parts of the bilinear products in Equation
(\ref{eq:wmatnew}) that relate to the new scattering contributions
in the extended theory. Since the new terms obey the special symmetry
that the initial and final substates are identical ($a=f$ and
$a^\prime =f^\prime$), there are reasons to assume full symmetry (time
reversal symmetry) between the absorption and emission legs of the
scattering transition. In this case the CRD branching ratio $B^{(K)}$
vanishes, and only the frequency coherent term with collisional
depolarization factor $A$ remains. These considerations lead us to the
following expression: 
\begin{eqnarray}
\label{eq:pprdnew} 
&{\cal P}_{\rm new} (\omega,\omega^\prime) =\textstyle{1\over
  2}[\Phi_{\,b-a}(\omega^\prime)+\Phi_{\,b-a^\prime}^\ast(\omega^\prime)]
\,k_g\,\cos^2\beta_g\,e^{2i\,\beta_g}\,\,A\,\delta(\omega-\omega^\prime) \,.
\end{eqnarray}
While $A$ is given by Equation (\ref{eq:brii}), the 
decoherenc angle $\beta_g$ is defined by Equation
(\ref{eq:betag}) if we let $\gamma$ be given by Equation
(\ref{eq:gamc}). Furthermore, for reasons explained in detail in Section
\ref{sec:nonzerosplit}, we need to apply the 
decoherence factor twice to the new interference terms, in contrast to
standard scattering theory. This is why the decoherence factor appears
in squared form in Equation (\ref{eq:pprdnew}), 

This expression leads to qualitatively
good agreement with the laboratory results for scattering of linear
polarization at potassium gas, but the predicted 
amplitudes of phase matrix elements $P_{21}$ and
$P_{22}$ are larger than the observed ones by a
factor of 2-3. For this reason we have in Equation (\ref{eq:pprdnew})
introduced an ad hoc scaling factor $k_g$, to be used as a free
parameter in the model fitting. The best fits are obtained with
$k_g\approx 0.4$. 

Physically a scaling factor $k_g$ that is less than unity implies that
the new interference terms, which arise from the splittings of the
ground states and which are the only source of the 
polarization effects that are observed for K D$_1$ scattering, are more
vulnerable to collisional destruction than accounted for by the
standard collisional depolarization factor $A$. The simplest way to
parametrize this enhanced collisional vulnerability is in terms of a
single scale factor.

\subsection{Incoherent Scattering}\label{sec:incoh}
The incoherent scattering matrix $\vect{R}_{\rm ic}$ represents the case
when collisions have erased the atomic polarization or ``memory'' of
the excitation process, making the emission process unrelated to the
absorption process. The expression for the normalized incoherent scattering
matrix is \citep[cf.][p.~80]{stenflo-book94} 
\begin{equation}
\vect{R}_{\rm ic}=[\,\vect{\Phi}(\omega)\,\vect{1}]\,[\vect{1}^T\,\vect{\Phi}(\omega^\prime)]\,,
\label{eq:rincexpl} 
\end{equation}
where $\vect{\Phi}$ here represents the $4\times 4$ Zeeman Mueller absorption
matrix without anomalous dispersion. It is normalized such that
integration of its first element, $\Phi_I$, 
over all frequencies is unity. In more explicit form $\vect{R}_{\rm
  ic}$ becomes  
\begin{equation}
\vect{R}_{\rm ic}=\begin{pmatrix}
\Phi_I \Phi_I^\prime &\Phi_I
\Phi_Q^\prime &\Phi_I \Phi_U^\prime
&\Phi_I \Phi_V^\prime \\ \Phi_Q
\Phi_I^\prime &\Phi_Q \Phi_Q^\prime
&\Phi_Q \Phi_U^\prime &\Phi_Q
\Phi_V^\prime \\ \Phi_U \Phi_I^\prime
&\Phi_U \Phi_Q^\prime &\Phi_U
\Phi_U^\prime &\Phi_U \Phi_V^\prime \\ \Phi_V \Phi_I^\prime &\Phi_V \Phi_Q^\prime &\Phi_V \Phi_U^\prime &\Phi_V \Phi_V^\prime \end{pmatrix}\,,
\label{eq:ric} 
\end{equation}
where  the primed functions $\Phi^\prime$ relate to the incident
frequency $\omega^\prime$, while the unprimed $\Phi$ relate to the
scattered frequency $\omega$. 

The explicit expressions for the functions $\Phi_{I,Q,U,V}$ are 
\begin{eqnarray}\label{eq:phiiquvexpl} 
\Phi_I &=&\Phi_\Delta \sin^2\gamma +{\textstyle{1\over 2}}
(\phi_+ +\phi_-)\,,\nonumber\\ \Phi_Q &=&\Phi_\Delta \sin^2\gamma\,\cos 2\chi
\,,\nonumber\\ \Phi_U &=&\Phi_\Delta \sin^2\gamma\,\sin 2\chi\,,\\ \Phi_V 
&=&{\textstyle{1\over 2}}(\phi_+-\phi_-)\cos\gamma\,,\nonumber\\ \Phi_\Delta 
&=&{\textstyle{1\over 2}}[\,\phi_0-{\textstyle{1\over 2}}
(\phi_+ +\phi_-)\,]\,.\nonumber
\end{eqnarray}
The $\phi_q$ profile functions in Equation
(\ref{eq:phiiquvexpl}) are obtained by summing over the real part of
the profile
functions of all transitions for which lower level $m$ minus 
upper level $m$ equals $q$, then weight these functions with the
respective transition probabilities (squares of the respective matrix
elements), and finally do Doppler broadening and area normalization.


\end{document}